\documentclass[%
 reprint,
superscriptaddress,
 amsmath,amssymb,
 aps,
]{revtex4-2}

\usepackage{graphicx}
\usepackage{dcolumn}
\usepackage{bm}

\usepackage[utf8]{inputenc}
\usepackage[english]{babel}
\usepackage[T1]{fontenc}
\usepackage{amsmath}
\usepackage{hyperref}

\usepackage{tikz}
\usepackage{lipsum}
\usepackage{xcolor}
\usepackage[ruled]{algorithm2e}
\usepackage{algpseudocode}
\usepackage{algpseudocode}

\usepackage{physics}
\usepackage{graphicx}
\usepackage{amssymb}

\DeclareMathOperator*{\argmin}{argmin}

\begin{document}

\preprint{BP-VQA-BL}

\title{Surviving The Barren Plateau in Variational Quantum Circuits \linebreak  with Bayesian Learning Initialization}

\author{Ali Rad}
\affiliation{
Joint Quantum Institute and Department of Physics,
University of Maryland, College Park, Maryland 20742, USA}

\author{Alireza Seif}
\affiliation{Pritzker School of Molecular Engineering, University of Chicago, Chicago, IL 60637}

\author{Norbert M. Linke}
\affiliation{
Joint Quantum Institute and Department of Physics,
University of Maryland, College Park, Maryland 20742, USA}
\affiliation{
Duke Quantum Center and Department of Physics, 
Duke University, Durham, North Carolina 27708, USA}

\date{\today}

\begin{abstract}
 Variational quantum-classical hybrid algorithms are seen as a promising strategy for solving practical problems on quantum computers in the near term. While this approach reduces the number of qubits and operations required from the quantum machine, it places a heavy load on a classical optimizer. While often under-appreciated, the latter is a computationally hard task due to the barren plateau phenomenon in parameterized quantum circuits. The absence of guiding features like gradients renders conventional optimization strategies ineffective as the number of qubits increases. Here, we introduce the fast-and-slow algorithm, which uses Bayesian Learning to identify a promising region in parameter space. This is used to initialize a fast local optimizer to  find the global optimum point efficiently. We illustrate the effectiveness of this method on the Bars-and-Stripes (BAS) quantum generative model, which has been studied on several quantum hardware platforms. 
Our results move variational quantum algorithms closer to their envisioned applications in quantum chemistry, combinatorial optimization, and
quantum simulation problems.

\end{abstract}

\maketitle


\section{\label{introduction}Introduction}

Quantum-classical hybrid algorithms are based on parameterized quantum circuits (PQC) that can prepare different quantum states through variable gate parameters. These algorithms can be adapted to different hardware environments and are, in principle, capable of solving a vast array of problems~\cite{cerezo2021variational}. This is achieved by outsourcing some of the computational complexity from the quantum device (QPU) to a classical processor (CPU). This keeps the quantum circuits shallow and amenable to noisy devices. 

These ideas have been used for developing variational quantum algorithms (VQAs)~\cite{mcclean2016theory}, such as the  Variational
Quantum Eigensolver (VQE)~\cite{peruzzo2014variational}, the Quantum
Approximate Optimization Algorithm (QAOA)~\cite{farhi2014quantum}, and Quantum Neural Network (QNN) architectures~ \cite{schuld2019quantum,schuld2014quest,wiebe2015quantum}. 
The original problem is mapped to finding the PQC parameters that minimize a cost function, which is evaluated by performing measurements on the circuit output of the QPU. The results are then provided to the CPU, which employs a classical optimization, or learning, algorithm to find the next set of parameters to feed back to the QPU in an iterative loop. 
Multiple demonstrations of this quantum-classical hybrid scheme have been realized on small systems~\cite{kandala2017hardware,peruzzo2014variational,dumitrescu2018cloud, hempel2018quantum,o2016scalable,kokail2019self,otterbach2017unsupervised}. The limitations are usually attributed to imperfect quantum hardware, but some of the work points out the importance of the CPU itself~\cite{zhu2019training}. 

The classical part of the algorithm is challenging for several reasons. The stochastic nature of QPU readout makes the measured cost function value fluctuate even for a fixed set of parameters. 
In addition, as the Hilbert space size and the parameter space size increase, the difficulty of finding the global minimum increases exponentially. This indicates that much like non-trivial NP-hard optimization problems, getting trapped in a local minimum is very likely~\cite{bittel2021training,larocca2021theory}.

Additionally, finding the optimal point is made even more difficult by a phenomenon called the "barren plateau" which means that far from any minima, the cost function provides no features to guide the optimization. It can arise for many circuit architectures, including ansatzes with a global cost function\cite{cerezo2021cost}, highly expressive ansatze circuits\cite{holmes2021connecting}, highly entangled \cite{patti2021entanglement,marrero2021entanglement} or noisy circuits\cite{wang2021noise,du2021learnability}, and the majority of  dissipative perceptron-based Quantum Neural Networks (QNN) \cite{sharma2020trainability}.  
It also provides a challenge for parameter initialization, since  random initialization of VQAs leads to exponentially small gradients \cite{mcclean2018barren}.

The existing optimization approaches can generally be divided into gradient-based and gradient-free methods.  In the former, the gradient information can be obtained via the parameter shift rule~
\cite{guerreschi2017practical,schuld2019evaluating,mari2021estimating}, or by directly measuring the first- or higher-order partial derivative on the quantum
hardware~\cite{cerezo2021higher,cerezo2020impact}. The optimization is then performed using algorithms such as Stochastic Gradient Decent (SGD)~\cite{sweke2020stochastic},
Quantum Natural Gradient Decent~\cite{stokes2020quantum,kubler2020adaptive}, meta learning~\cite{wilson2021optimizing}, and Simultaneous Perturbation Stochastic Approximation
(SPSA)~\cite{spall1992multivariate}. The latter only uses the value of the cost-function and includes methods such as Nelder-Mead, COBYLA, Powell's, and Bayesian based methods~\cite{powell1994direct,powell1964efficient,arrasmith2021effect,zhu2019training,bonet2021performance,verdon2019learning}.
Experiments  have revealed the vulnerability of both gradient-based and gradient-free methods to barren plateaus~\cite{cerezo2021higher,arrasmith2021effect,mcclean2018barren,cerezo2021cost,sharma2020trainability,wang2021noise,holmes2021barren,zhang2020toward,marrero2021entanglement,patti2021entanglement,uvarov2021barren}. While strategies to mitigate or avoid the barren plateau have been proposed~\cite{larocca2021theory,uvarov2021barren,cerezo2021cost,holmes2021connecting,grant2019initialization,volkoff2021large,larocca2021theory}, their efficiency in general scenarios remains untested. Others require the circuit to be over-parmatrized \cite{larocca2021theory}, which might not be feasible, or demand exponentially scaling resources\cite{holmes2021barren}. 

In this paper, we introduce and evaluate a new technique, for finding the global optimum in VQAs, which we call fast-and-slow following Ref.~\cite{mcleod2018optimization}. It employs Bayesian learning, which is gradient-free, as an initialization procedure for subsequent gradient-based optimization, combining global and local information of the parameter landscape.  We first describe this method and then test it for different local optimizers on the Bars-and-Stripes (BAS) quantum generative model~\cite{benedetti2019generative,zhu2019training}. BAS is a quantum machine learning algorithm, which can be used as a benchmark to study the performance and capabilities of PQCs, and which shows barren plateaus.


\begin{figure}\label{fig1}
\includegraphics[width=0.5\textwidth]{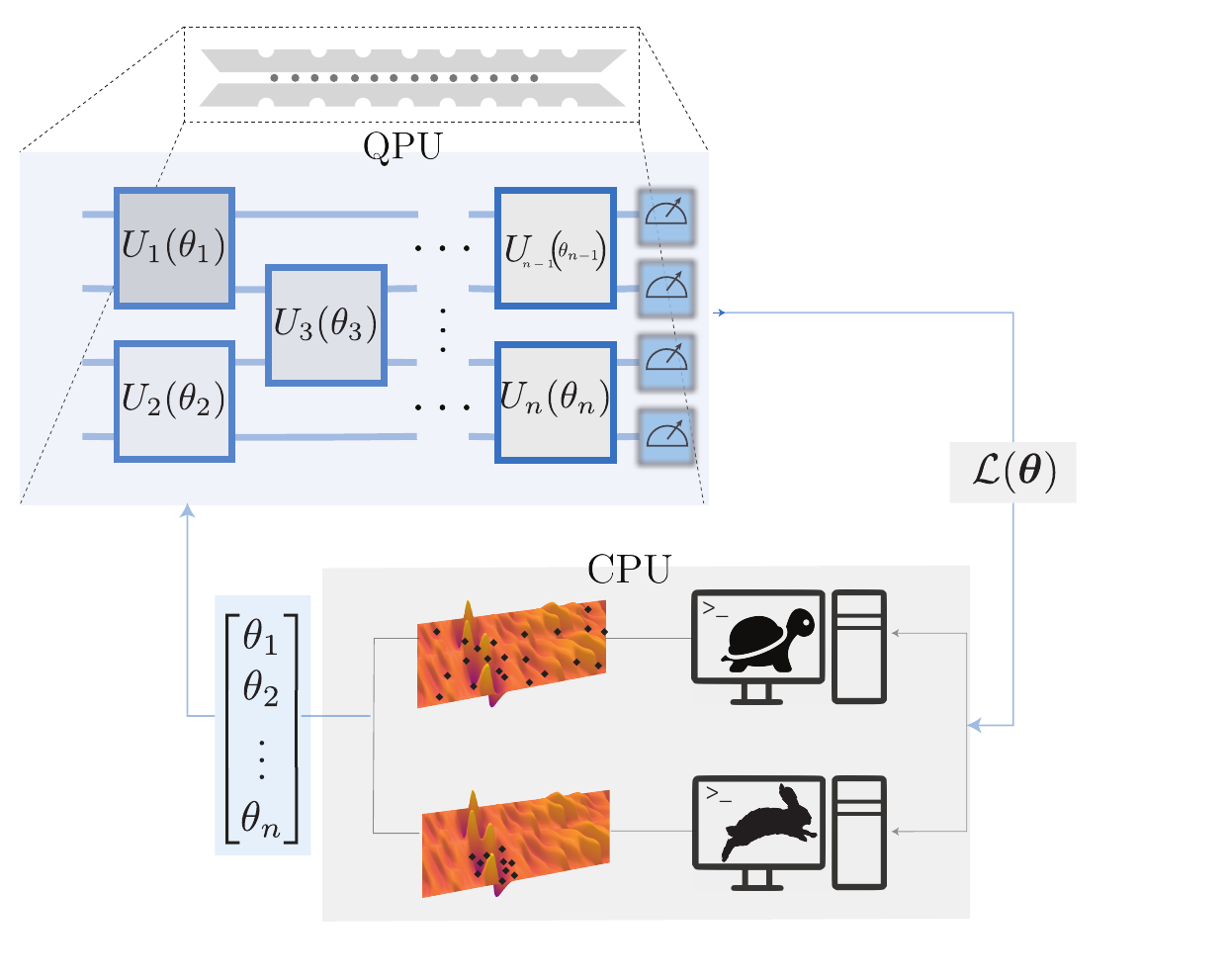}
\centering
\caption{The fast-and-slow optimization method for variational quantum algorithms. The QPU executes a quantum circuit consisting of gates $\{U(\theta_i)\}$, parameterized by $\boldsymbol\theta=\{\theta_i\}$. The output is used to calculate a cost function value $\mathcal{L}(\boldsymbol\theta)$, which is passed to an optimizer running on a CPU. The CPU first uses a slow global search method to identify a promising region in parameter space and then a fast local optimizer to find the minimum.}  
\end{figure}

\section{The Fast-and-Slow Method}\label{section_F_and_S}


Our method has two parts and is shown schematically in  Fig.\ref{fig1}. 
In the first, slow, part we initialize the parameters to zero and perform Bayesian optimization using Gaussian processes.
This method is well-suited for this task, since querying the QPU is expensive and results in  noisy outputs~\cite{frazier2018tutorial}. 
The computational complexity of Bayesian optimization increases as the number of samples $n$ gathered from the QPU accumulates, due to the $\mathcal{O}(n^3)$ scaling of the calculation of the co-variance matrix inverse ~\cite{shahriari2015taking}. Therefore,
this method is not suited for a detailed local search.



In the second, fast, part we use the best parameter set from the slow part to initialize a local optimizer. This is now highly likely to reach the global optimum, since we start in the correct region and there is no longer a barren plateau~\cite{liu2021representation}. 

There is a trade-off between the number of queries devoted to the global and local optimizers. Too many queries in the slow part waste resources that should be spent on local optimization, while too few queries increase the chance to switch over in a region containing only a local minimum. The latter might lead to failure since there is no guarantee for a local optimizer to converge to the global optimum after random initialization, regardless of the number of iterations \cite{barton1991modifications}. In practice, the BO shows a distinctive cost function drop after a certain number of iterations for a given circuit. We use this phenomenological criterion as the change-over point in our protocol (see below). Additionally, around this point, the standard deviation of the cost function calculated for multiple batches of the experiment decreases significantly. Note that in general, we expect the switching point to be problem-dependent. The code is for the fast-and-slow algorithm used for this work available on GitHub~\footnote{\texttt{Fast and Slow Algorithm} codebase: \url{https://github.com/frustea/Quantum-Fast-and-Slow}.}.


 \section{Results}
\begin{figure*}[t]\label{output22}
\includegraphics[width=\textwidth]{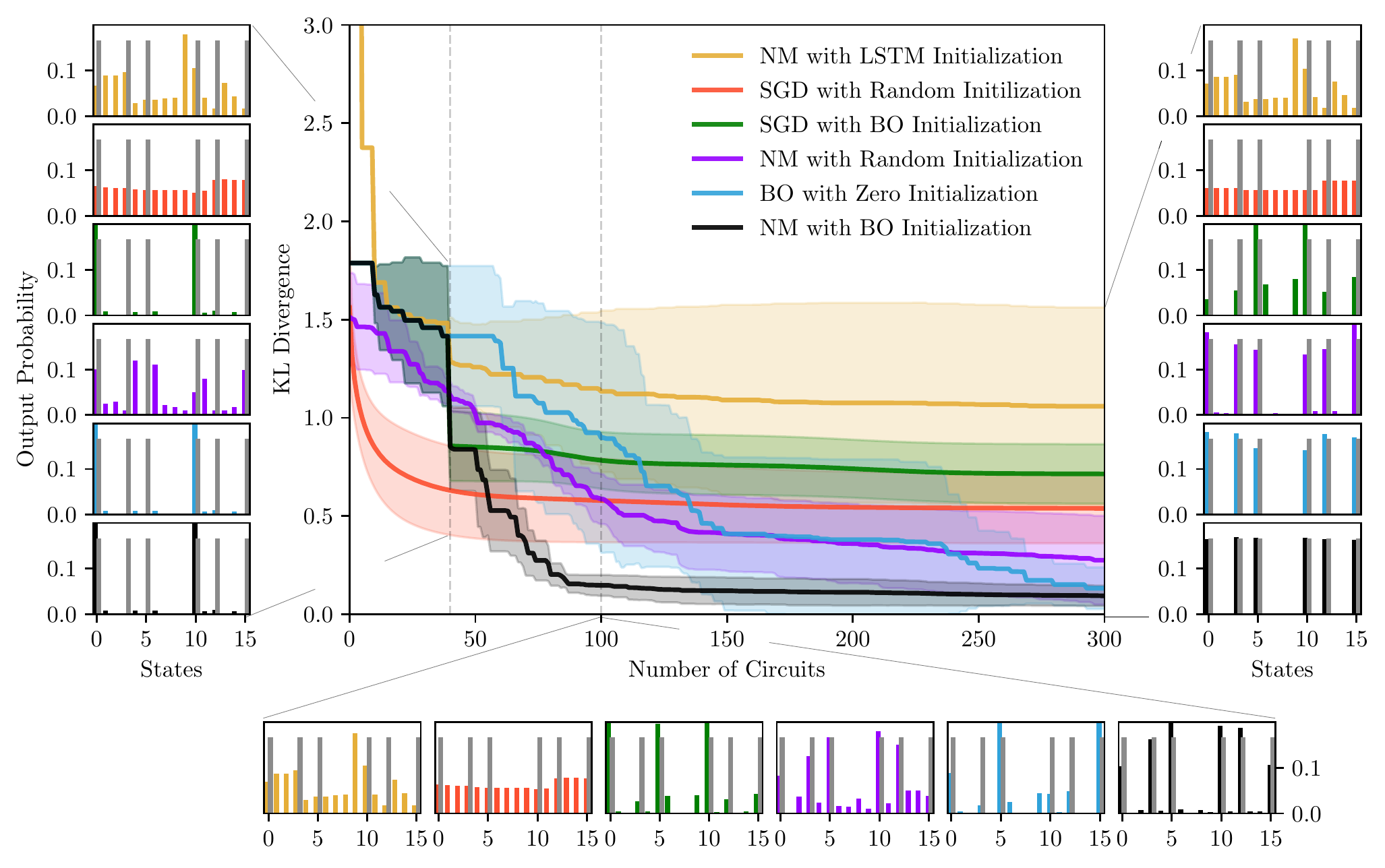}
\centering
\caption{Comparison of different optimization methods for the $\text{BAS}(2,2)$ problem with 4 qubits and 26 circuit parameters.   
The main graph shows the cost function~\eqref{KL_div} against the number of executed circuit instances on the simulated QPU for different combinations of initialization and optimization methods: Nelder-Mead (NM), Stochastic Gradient Descent (SGD), Neural Network (LSTM),  and Bayesian Optimization (BO). The fast-and-slow method corresponds to "NM with BO Initialization." For each method, the average value (solid line) and standard deviation based on five repetitions (shaded region) are shown. The output distribution is sampled $N_s=1024$ times to add statistical errors. The insets show the output distribution at three stages of optimization: 1) after 45 iterations (left), which is the switching point from BO to NM in the fast-and-slow method, 2) after 100 iterations (bottom), when the cost function of fast-and-slow is plateauing, and 3) after 300 iterations (right).}
\end{figure*}

\begin{figure*}\label{output23}
\includegraphics[width=\textwidth]{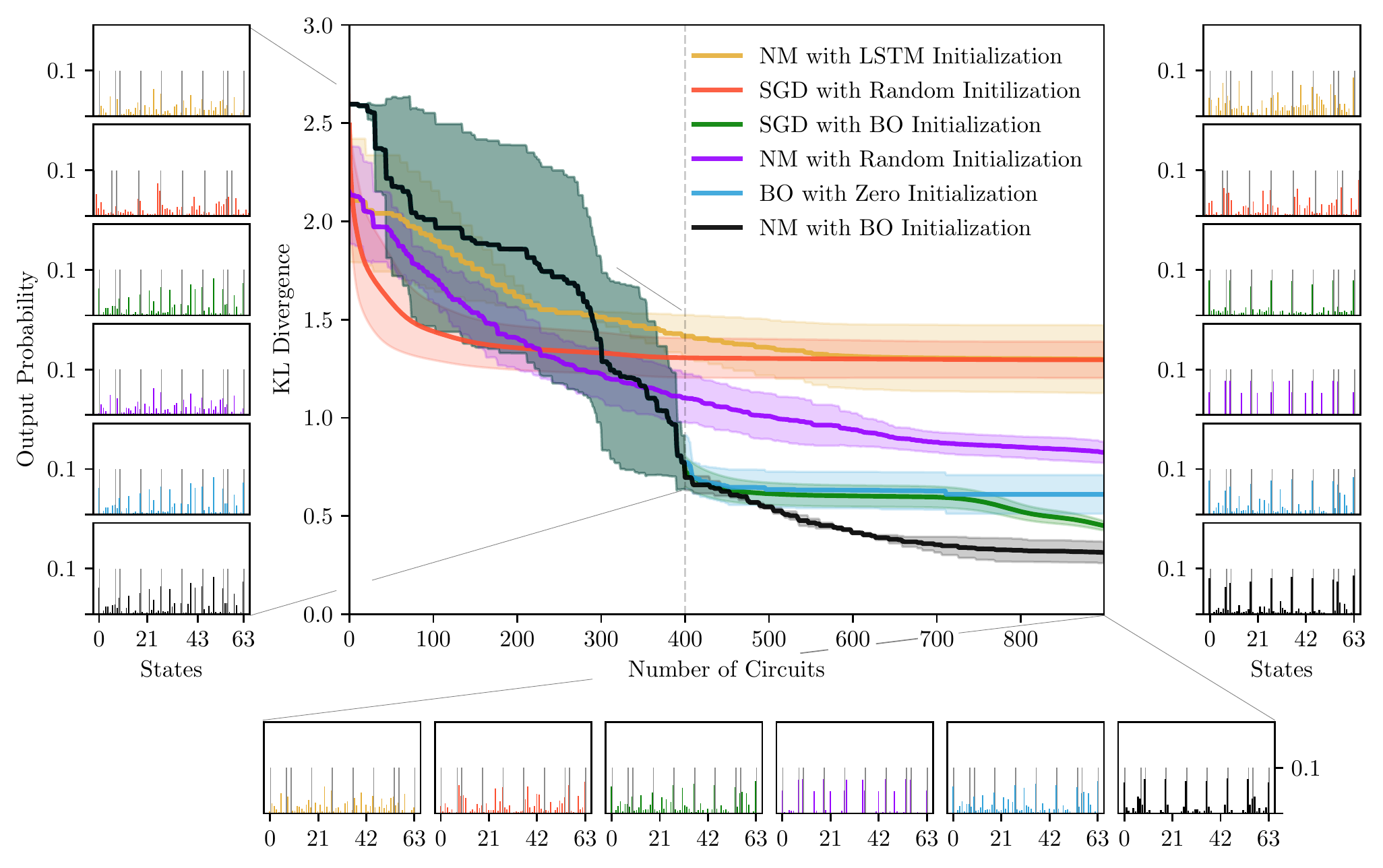}
\centering
\caption{ 
Comparison of different optimization methods for the $\text{BAS}(2,3)$ problem with six qubits and 41 circuit parameters (see text): Nelder-Mead (NM), Stochastic Gradient Descent (SGD), Neural Network (LSTM),  and Bayesian Optimization (BO). The fast-and-slow method "NM with BO Initialization" is still the best strategy and produces the correct output distribution, but the convergence is slower than for the four-qubit case. The insets show the output distribution at three stages of optimization: 1) after 400 iterations (left), which is the switching point from BO to NM in the fast-and-slow method, 2) after 900 iterations (bottom), and 3) after 1800 iterations (right), which is beyond the range shown in the main plot. 
} 
\end{figure*}





To evaluate or method, we simulate the optimization of BAS circuits on four and six qubits on a classical computer. The BAS \cite{mackay2003information} maps qubit states in the computational basis to a two-dimensional array of  black or white pixels, see appendix \ref{appendixB} for details. This problem is a good test case since its convergence behavior has been studied on a trapped ion system and it was found to be very well captured by a simple finite-sampling noise model \cite{zhu2019training}.



Specifically, we consider the ensembles $\text{BAS}(2,2)$ and $\text{BAS}(2,3)$ with input $|\psi\rangle=|0\rangle^{\otimes n}$ for $n=2\times 2$ and $n=2\times 3$, respectively. The circuit design follows \cite{zhu2019training} with qubits connectivity given by star-graph, and uses a gate set native to trapped ions. The first layer consists of single-qubit $X$ and $Z$ rotation operators. The following layer applies $XX$ entangling gates to all pairs of qubits. The unitary operator associated with the mentioned circuits, for a total number of layers $L$, can be written as:
\begin{equation}
     U(\boldsymbol{\theta})=\prod_{k=1}^{|\boldsymbol{\theta}|} U_k(\theta_k)=\prod_{\ell=1}^L\prod_{\substack{i=1\\j=2}}^{n} R_i(\alpha^\ell_i,\beta^\ell_i,\gamma^\ell_i)R_{x_1 x_j}({\phi}_{j}^\ell)
 \end{equation}
where we explicitly identify $U_k$ and $\theta_k$ corresponding to one and two qubit gates and their parameters in layer $\ell$ with 
 \begin{align}
 R_{x_i x_j}(\phi_{j}^\ell)&=e^{-i \phi_{j}^\ell X_i X_j}\\
     R_i(\alpha^\ell_i,\beta^\ell_i,\gamma^\ell_i)&= e^{i\alpha^\ell_iX_i}e^{i\beta^\ell_iZ_i}e^{i \gamma^\ell_iX_i}.
 \end{align}
Expressed in this gate set, the quantum circuits for BAS(2,2) and BAS(2,3) have 26 and 41 variational parameters, respectively.


 To ease comparison with \cite{zhu2019training}, the Kullback–Leibler (KL) divergence is used as the cost function, which is a standard metric to compare two distributions \cite{kullback1951information}:
\begin{equation}\label{KL_div}
    \mathcal{L}_{\rm{KL}}(\boldsymbol\theta)=\text{KL}\Big(\Tr(OU^\dagger(\boldsymbol\theta)\rho U(\boldsymbol\theta))||\rho_{\text{BAS}}\Big).
\end{equation}

Since the variance of the cost function gradient, $\text{Var}[\partial_\mu \mathcal{L}]=\langle (\partial_\mu \mathcal{L})^2- \langle \partial_\mu \mathcal{L}\rangle^2 \rangle$, is exponentially suppressed as a function of qubit number,

\begin{equation}
    \text{Var}[\partial_\mu \mathcal{L}]\leq \frac{1}{2^{6n}}f(O,\rho,\rho_{\text{BAS}})
\end{equation}
there exists a barren plateau in the optimization landscape (see appendix \ref{appendix_A}).

To study the performance of the fast-and-slow method, we compare it against strategies with only local or only global optimizers. Specifically, for the pure local schemes, we use the Stochastic Gradient Decent (SGD) \cite{shalev2014understanding,harrow2021low,sweke2020stochastic}  and Nelder-Mead (NM) algorithm~\cite{nelder1965simplex} with random initialization. As a global scheme we consider BO with parameters initialized to zero. For the fast-and-slow method, we employ the slow BO phase as discussed above, followed by either NM or SGD. Additionally, we consider another hybrid algorithm introduced in Ref.~\cite{verdon2019learning}, that utilizes a Long Short Term Memory (LSTM) recurrent neural network for initialization followed by NM.

The results are shown for four qubits in Fig.~\ref{output22} and six qubits in Fig.~\ref{output23}. The effectiveness of a classical optimizer for VQAs is typically only assessed by the convergence rate of the cost-function without taking account the quality of the output circuit~\cite{zhou2020quantum,grant2019initialization,wiersema2020exploring,skolik2021layerwise,campos2021abrupt,verdon2019learning}. However, this does not in general make clear when the algorithm gets trapped in a local minimum~\cite{hamilton2019generative}. For comparison with the target distributions, we show the state populations at different stages of training as figure insets.

The results show that the fast-and-slow algorithm (NM with BO initialization) outperforms the other methods based on the convergence behavior, the final value of the cost function, and its success in  generating  the desired output distribution. This is even more pronounced when the number of qubits is increased to six, shown in Fig.~\ref{output23}. 

In our investigations, pure NM was found to perform better than other standard local gradient-free optimizers on their own, such as COBYLA~\cite{powell1994direct} and Powell~\cite{powell1964efficient}. Nevertheless, its failure to find the true minimum for six qubits confirms that simplex-based gradient-free optimizers are as susceptible to local minima as gradient-based methods~\cite{barton1991modifications}.

The fast-and-slow variant involving BO initialization followed by the gradient-based SGD method is less effective than using NM as the fast stage. The output distributions only partially match the ideal ones, especially when there are more qubits involved. Additionally, pure BO has a low convergence rate and a large classical computational overhead, which means that despite the reliability of discovering the global optimum, our results confirm that it is not practical for VQAs. Finally, LSTM is the weakest strategy for solving our benchmark problems.

\

\section{OUTLOOK}

Our results show that the fast-and-slow method of initialization and optimization is highly promising since it reduces the number of queries to the QPU substantially and makes training the VQA more practical for a larger number of qubits. The simulated experiment, while well-motivated by being a generic circuit and allowing the comparison with a recent experimental implantation on quantum hardware, only represents a single example problem. Going forward, we will evaluate the method on different kinds of circuits on physical, rather than simulated, quantum hardware. 

Furthermore, the fast-and-slow algorithm introduced here represents only the simplest form of a combined scheme with a single switching point and fixed switching criterion. More complex problems might require a dynamical alternation between the fast and slow parts, which has been shown to be beneficial for some problems in classical optimization~\cite{mcleod2018optimization}.
Further study is needed to find appropriate methods for determining the switching point and problem-specific adaptations, which will make this method even more powerful.


\begin{acknowledgments}
We are extremely grateful for the support of Mind Foundry for providing access to their  OPTaaS Bayeisan optimizer \cite{OPTaaS}.  N.M.L. acknowledges funding by the Office of Naval Research (N00014-20-1-2695) and the Maryland-Army-Research-Lab Quantum Partnership (W911NF1920181). A.S. is supported by a Chicago Prize Postdoctoral Fellowship in Theoretical Quantum Science. This work received support from the National Science Foundation through the Quantum Leap Challenge Institute for Robust Quantum Simulation (OMA-2120757).
\end{acknowledgments}

\appendix

\section{The Barren Plateau in VQAs}\label{appendix_A}
Variational circuits can be described as a unitary operation, $U(\boldsymbol\theta)$, with a set of parameters  $\boldsymbol\theta=\{\theta_i\}_{i=1}^m$
For a given observable $\mathcal{O}_i$, and a fixed initial state $\rho$, 
the expectation value 
can be estimated by executing  repetitive measurements on the QPU. In general, these measurements can be used to calculate the cost function defined as
\begin{equation}
   \mathcal{L}(\boldsymbol{\theta})= \sum_{i} h_i f_i\Big(\Tr\big(\mathcal{O}_iU(\boldsymbol{\theta} )\rho U^\dagger(\boldsymbol{\theta}))\Big),
\end{equation}
where $\{h_i\in \mathbb{R},f_i:\mathbb{R}\rightarrow \mathbb{R}\}$ encodes the  problem.  

 The goal of VQAs is to find the optimum set of parameters that minimizes the cost function:
\begin{equation}
    \boldsymbol\theta^{\star} = \argmin_{\boldsymbol\theta}\mathcal{L}(\boldsymbol\theta).
\end{equation}

 In general, the anzatz can be described with a unitary $U(\boldsymbol{\theta})$ as
 \begin{equation}
     U(\boldsymbol{\theta})=\prod_{i=1}^{m}U_i(\theta_i),
 \end{equation}
where  $U(\theta_i)=e^{-i\theta_i \sigma_i}$ and $\sigma_j$ is a Hermitian  1- or 2-qubit operator  and $\sigma_j^2=\mathbb{I}$. 
In this representation we assume the parameters are independent of each other. To study behaviour of a specific parameter $\theta_\mu$, $1\leq\mu\leq m$, we can split the $U(\boldsymbol{\theta})$ into a left and right part:
 \begin{equation}
     U(\boldsymbol{\theta})=U_R(\boldsymbol{\theta}_R)U_L(\boldsymbol{\theta}_L),
 \end{equation}

 where right and left operators are defined as $U_R(\boldsymbol{\theta}_R)=\prod_{i=1}^{\mu} U_i(\theta_i)$ and  $U_L(\boldsymbol{\theta}_L)=\prod_{i=\mu+1}^{m} U_i(\theta_i)$, respectively.   The derivative of $U(\boldsymbol\theta)$ can be written 
 as $\partial_{\theta_\mu} U(\boldsymbol\theta)= U_R (-\frac{i}{2}\sigma_{\mu })U_L$
 and $\partial_{\theta_\mu}U^{\dagger}(\boldsymbol\theta)=U_L^\dagger(\frac{i}{2}\sigma_{\mu}) U_R^\dagger$

The cost function of our generative model is based on the 
Kullback–Leibler (KL) divergence function in  Eq.~\eqref{KL_div} in the main text.
We define $C_i:= \text{Tr}(O_iU\rho U^\dagger)$, $q_i:=\Tr[O_i \rho_{\text{BAS}}]$, and let $O_i \in\{\ketbra{i}\}_{i=0}^{2^n-1}$. Note that  $ \partial_{\theta_\mu} \sum_i C_i=0$. Therefore, the derivative of the cost-function with respect to a single parameter, $\theta_\mu$, is given by
\begin{equation}\label{muL}
    \partial_{\mu}\mathcal{L}(\theta)=\sum_{i=0}^{2^n-1}\partial_{\theta_\mu} C_i \log (C_i/q_i),
\end{equation}
where the derivative of $C_i$ respect to the parameter $\theta_\mu$ can be written as ~\cite{cerezo2021cost}
 \begin{align}
\begin{split}
 \partial_\mu C_i&
    =-\frac{i}{2}\text{Tr}(U_R^\dagger O_i U_R [\sigma^\mu, U_L \rho U_L^\dagger])\\
    &=\frac{i}{2}\text{Tr}(U_L \rho U_L^\dagger [\sigma^\mu, U_R^\dagger O_i U_R]).
\end{split}
\end{align}

 In order to calculate the  average value of the gradient, $\langle  \partial_{\theta_k}\mathcal{L}(\theta) \rangle=\sum_i \langle \partial_\mu C_i\log C_i/q_i\rangle $, we can consider three cases: 
 either $U_L$, $U_R$, or both satisfy the 2-design property. This allows us to simplify certain terms in the average by an integral over the Haar measure.  A $t$-design unitary defined as a finite set of unitary operator$\{U_k\}_{k=1}^k \in \mathcal{U}(d)$ with any arbitrary function $P_{(t,t)}(U)$(which acts at each element of matrix $U$ and $U^\dagger$ with polynomial degree at most $t$) such that it satisfies the following relation
 \begin{equation}
     \frac{1}{K}\sum_{k=1}^K P_{(t,t)}(U_k)=\int_{\mathcal{U}(d)}d\mu(U) P_{(t,t)}(U),
 \end{equation}
where $d\mu(\cdot)$ is a Haar measure over the unitary group $\mathcal{U}(d)$.
 In general, for a random unitary matrix $U=(U_{ij})_{i\leq i,j\leq d}$, the expectation of the following form with respect to the Haar measure over a unitary group $\mathcal{U}(d)$ is given by
 \begin{align}\label{weingarten}
    \begin{split}
        &\mathbb{E}[U_{i_1 j_1} U_{i_2 j_2} \cdots U_{i_n j_n}U^*_{i'_1 j'_1}U^*_{i'_2 j'_2}\cdots U^*_{i'_n j'_n}]\\
        &=\sum_{\sigma, \tau \in S_n}\delta_{i_1i'_\sigma(1)}\cdots \delta_{i_n i'_{\sigma(n)}}\delta_{j_1j'_\sigma(1)}\cdots \delta_{j_n j'_{\sigma(n)}}\text{Wg}^U(\sigma^{-1}\tau,d)\\
        &:=\sum_{\sigma, \tau \in S_n}\delta_\sigma(\bold{i},\bold{i}')\delta_\tau(\bold{j},\bold{j}')\text{Wg}^U(\sigma^{-1}\tau,d),
    \end{split}
\end{align}

where $S_n$ is the symmetric group  ~\cite{8178732,puchala2011symbolic}. The  $\text{Wg}(\cdot,d)$ is Weingarten function defined on $S_d$ with the following Fourier expansion:
\begin{equation}
    \text{Wg}^U(\sigma,d)=\frac{1}{d!}\sum _{\lambda \vdash n , \ell(\lambda)\leq d}\frac{f^\lambda}{\prod_{i=1}^{\ell(\lambda)}\prod_{j=1}^{\lambda_i}(d+j-1)}\chi^{\lambda}(\sigma),
\end{equation}
such that for a given $\lambda$, $\chi^\lambda$ is the irreducible character of $S_d$ associated with it. $f^\lambda$ is the degree of $\chi^\lambda$ and 
 $\lambda \vdash n $ is the sum over all partitions of $\lambda$ defined by Young diagram $\lambda=(\lambda_1,\lambda_2,\cdots \lambda_l)$ of $d$ and $l=\ell(\lambda)$. Applying this theorem for first and second moment of  $U$, over the Haar measure $d\mu(U)$, provides the following identities based on the calculation over  $S_1$ and $S_2$ group: 
 \begin{align}\label{identity_1}
    &\int U_{ij}U_{pk}^* d\mu(U)=\frac{\delta_{ip}\delta_{jk}}{d}\\
    &\int U_{i_1j_1}U_{i_2j_2}U^*_{i'_1 j'_1}U^*_{i'_2,j'_2}d\mu(U)=\frac{\delta_{i_1i'_1}\delta_{i_2i'_2}\delta_{j_1j'_1}\delta_{j_2j'_2}}{d^2-1}\nonumber \\
    &-\frac{\delta_{i_1i'_1}\delta_{i_2i'_2}\delta_{j_1j'_2}\delta_{j_2j'_1}+\delta_{i_1i'_2}\delta_{i_1i'_2}\delta_{j_1j'_1}\delta_{j_2j'_2}}{d(d^2-1)} \nonumber.
\end{align}
For $U\in \mathcal{U}(d=2^n)$ the above identities  can  be re-written in the form of
\begin{align}\label{identity_2}
    &\int \text{Tr}(UXU^\dagger Y)d\mu(U)=\frac{1}{2^n}\text{Tr}(X)\text{Tr}(Y)\\
    &\int \text{Tr}(UXU^\dagger Y)\text{Tr}(UZU^\dagger W)d\mu(U)= \nonumber \\
    &\frac{\Tr(X)\Tr(Y)\Tr(Z)\Tr(W)+\Tr(XZ)\Tr(YW)}{2^{2n}-1} \nonumber \\
    &-\frac{\Tr(XZ)\Tr(Y)\Tr(W)+\Tr(X)\Tr(Z)\Tr(YW)}{2^{3n}-2^n} \nonumber,
\end{align}
and
\begin{align}\label{identity_3}
    &\int \Tr[UXU^\dagger Y U Z U^\dagger W]d\mu(U)\\
    &=\frac{\Tr(X)\Tr(Z)\Tr(YW)+\Tr(XZ)\Tr(Y)\Tr(W)}{2^{2n}-1}\nonumber \\
    &-\frac{\Tr(XZ)\Tr(YW)+\Tr(X)\Tr(Y)\Tr(Z)\Tr(W)}{2^n(2^{2n}-1)}.
\end{align}
For simplicity,we start the calculations with the case that both $U_R$ and $U_L$ are 2-designs. By applying the above identities to calculate $\mathbb{E}[C_i]$ and $\mathbb{E}[C_iC_j]$, we obtain
\begin{align}\label{Eq:E[C]}
    \mathbb{E}[C_i]&=\int \Tr(O_i U\rho U^\dagger) d\mu(U)\\
    &=\frac{\Tr(O_i)\Tr(\rho)}{2^n}= \frac{1}{2^n}\nonumber,
\end{align}
and
\begin{align}\label{Eq:E[C^2]}
\begin{split}
    \mathbb{E}[C_i C_j]&=\int \Tr(O_i U\rho U^\dagger)\Tr(O_j U\rho U^\dagger) d\mu(U)\\
    &=\frac{\Tr(O_i)\Tr(O_j)}{2^{2n-1}}[\Tr(\rho)^2-\frac{\Tr(\rho^2)}{2^n}]\\
    &+\frac{\Tr(O_i O_j)}{2^{2n}-1}[\Tr(\rho^2)-\frac{\Tr(\rho)^2}{2^n}]\\
    &=\frac{1+\delta_{i,j}}{2^{2n}-1}(1-{2^{-n}}),
\end{split}
\end{align}
since for observable in the form of  $O_i^{jk}=\delta_{i,j}\delta_{i,k}I_{2^n\times 2^n}^{jk}$, we have $\Tr(O_i)=\Tr(O_i^2)=1$ and $\sum_{i=0}^{2^{n-1}}\Tr(O_i)=2^n$. A similar calculation can be repeated for the $\mathbb{E}[\partial_\mu C_i]$:
\begin{align}\label{E[partial C]}
\begin{split}
    \mathbb{E}[\partial_\mu C_i]&=\frac{i}{2}\int \Tr(U_L\rho U_L^\dagger[\sigma^\mu, U_R^\dagger O_iU_R])d\mu(U_L)d(U_R)\\
    &=i\frac{\Tr(\rho)[\Tr(\sigma^\mu),\Tr(O_i)]}{2^{2n+1}}=0,
\end{split}
\end{align}
and therefore $\mathbb{E}[\partial_\mu C_i \log q_i]=0$. Similarly the expectation value of the second moment  can be calculated as a follows: 


\begin{align}\label{Eq:E[(partial C)^2}
\begin{split}
    \mathbb{E}[\partial_\mu C_i\partial_\mu C_j]&= -\frac{1}{4}\int \Tr(U_L\rho U_L^\dagger[\sigma^\mu, U_R^\dagger O_iU_R])\\
    \times&\Tr(U_L\rho U_L^\dagger[\sigma^\mu, U_R^\dagger O_j U_R])d\mu(U_L,U_R)\\
     =& \Big(\frac{\Tr(\sigma_\mu)^2 \Tr(O_iO_j)+\Tr(\sigma_\mu^2)\Tr(O_i)\Tr(O_j)}{2^{2n}-1}\\
       -&\frac{\Tr(\sigma_\mu^2)\Tr(O_iO_j)+\Tr(\sigma_\mu)^2
    \Tr(O_i)\Tr(O_j)}{2^{3n}-2^n}\\
        -&\frac{1}{2^n}\Tr(\sigma_\mu^2)\Tr(O_iO_j)\Big)\times(\frac{-\Tr(\rho^2)+2^{-n}}{2^{2n}-1}).
\end{split}
\end{align}

To obtain an upper-bound on $|\langle \partial_\mu \mathcal{L}\rangle |$ we use the Cauchy-Schwarz inequality and the property of logarithmic functions, where for a two random variable $x$ and $y$,   we have $\mathbb{E}[x\log y]\leq \mathbb{E}[xy]-\mathbb{E}[x]$. Applying this identity to calculate $\mathbb{E}[\partial_\mu C_i {C_i}/{q_i}]$ leads to the following inequality:
\begin{align}
\begin{split}
\Big|\mathbb{E}[\partial_\mu C_i \log \frac{C_i}{q_i}]\Big|&\leq \Big|\mathbb{E}[\partial_\mu C_i \frac{C_i}{q_i}]-\mathbb{E}[\partial_\mu C_i]\Big|=\Big|\mathbb{E}[\partial_\mu C_i \frac{C_i}{q_i}]\Big|\\
&\leq q_i^{-1}\sqrt{\mathbb{E}[(\partial_\mu C_i)^2]}\sqrt{\mathbb{E}[(C_i)^2]}.
\end{split}
\end{align}


To avoid the divergence in the numerical calculation of KL divergence we replace  the $q_{i,\text{BAS}}$ with the clipper function $\max[\epsilon, q_{i,\text{BAS}}]$ for arbitrary small $\epsilon:=\Lambda^{-1}$. By using  the result of Eq~.\eqref{Eq:E[(partial C)^2} and Eq.~\eqref{Eq:E[C]} and adding all terms from Eq~.\eqref{muL}, we can derive the following upper bound on the magnitude of  expectation value gradient of the cost function: 

\begin{equation}
    |\langle \partial_\mu \mathcal{L}\rangle |\leq \sqrt{\frac{2(2^{n}-1)^2(\Lambda\cdot 2^n+2\Lambda-4)^2}{2^{2n}(2^n+1)(2^{2n}-1)^2}}\sim \mathcal{O}( 2^{-\frac{3n}{2}}).
\end{equation}

In the next step, to find the expectation value of the variance of the gradient of the cost function with respect to a parameter $\theta_\mu$, $\text{Var}[\partial_\mu \mathcal{L}]$, we start with the its definition and use Cauchy-Schwarz to get an upper bound on it 
\begin{align}
\begin{split}\label{Var C}
    \text{Var}[\partial_\mu \mathcal{L}]&=\langle (\partial_\mu \mathcal{L})^2- \langle \partial_\mu \mathcal{L}\rangle^2 \rangle\\
    &=\sum_{i, j}\mathbb{E}[ \partial_\mu C_i \log \frac{C_i}{q_i}  \partial_\mu C_j \log \frac{C_j}{q_j} ]\\
    &-\sum_i\mathbb{E}[\partial_\mu C_i \log \frac{C_i}{q_i}]^2\\
    &\leq \sum_{i, j}\mathbb{E}[ \partial_\mu C_i \frac{C_i}{q_i}  \partial_\mu C_j \frac{C_j}{q_j} ]\\
    & \leq \sum_{i,j}(q_iq_j)^{-1}\Big( \mathbb{E}[(\partial_\mu C_i)^4]\mathbb{E}[(\partial_\mu C_j)^4]\mathbb{E}[C_i^4]\mathbb{E}[C_j^4]\Big)^{\frac{1}{4}}.
\end{split}
\end{align}
To calculate the fourth moment expressions in the right-hand side of the inequality, we assume the both $U_R$ and $U_L$ satisfy the 4-design property. This assumption is motivated by the observation that random local quantum circuits with a depth that grows polynomially  in the number of quibits form an approximate unitary $t$-design, and conjectured to be valid for logarithmic-depth circuits as well  ~\cite{brandao2016local,marrero2021entanglement,harrow2018approximate}. Based on the  4-design assumption,  according to Eq.~\eqref{weingarten}:

 \begin{align}\label{C^4}
     \begin{split}
         \mathbb{E}[C_i^4]&=\sum_{\bold{i,i',j,j'}=1}^4\mathbb{E}[U_{i_1 j_1}U_{i_2 j_2}U_{i_3 j_3}U_{i_4 j_4}U^*_{i'_1 j'_1}U^*_{i'_2 j'_2}U^*_{i'_3 j'_3}U^*_{i'_4 j'_4}]\\
         &\times \prod_{\nu=1}^4 \rho_{j_\nu i'_\nu} O^{i}_{j'_\nu i_\nu}=\sum_{\substack{i_1,j_1, \cdots ,i_4,j_4\\ i'_1,j'_1,\cdots ,i'_4,j'_4=1}}^4 \sum_{\sigma, \tau \in S_4}\prod_{\nu=1}^4\\
         &\times 
       \rho_{j_\nu i'_\nu} O^{i}_{j'_\nu i_\nu}\delta_{\sigma}(\bold{i},\bold{i}')\delta_\tau(\bold{j},\bold{j}')\text{Wg}^U(\sigma^{-1}\tau,d)\\
       &=\sum_{\sigma,\tau \in S_4}\sum_{\substack{i'_1,\cdots i'_4\\ j'_1,\cdots j'_4=1}}^4\prod_\nu^4 O^i_{i'_{\sigma(\nu) j'_\nu }}\rho_{j'_{\tau(\nu)}i'_\nu} \text{Wg}^U(\sigma^{-1}\tau,d)\\
       &:=\sum_{\sigma,\tau \in S_4}\Tr_{\sigma, \tau}(\rho, O^i)\text{Wg}^U(\sigma^{-1}\tau,d)\\
       &\leq \frac{C}{d^4}\sum_{\sigma,\tau\in S_4}\Tr_{\sigma, \tau}(\rho, O^i),
     \end{split}
 \end{align}
where in  last inequality  based on the property of Weingarten coefficient ~\cite{roth2018recovering} and C is a constant number. Similarly 
 \begin{align}\label{muC^4}
     \begin{split}
        \mathbb{E}[(\partial_\mu C_i)^4]&=\sum_{\substack{\sigma, \tau \in S_4\\
        \alpha, \beta}} \text{Wg}^U_L(\sigma^{-1}\tau,d)\text{Wg}^U_R(\alpha^{-1}\beta,d)\sum_{\substack{\bold{i',j',k}\\ \bold{k',l',l}=1}}^4  \\ 
        \sum_{\substack{r_v: r_1 r_2\\r_3,r_4=0}}^1&\prod_{\nu=1}^4
        \times {(\rho_{j'_{\tau(\nu)}i'_\nu}\sigma^\mu_{j'_\nu k_\nu}O^i_{l'_\nu k_\nu}\delta_{k_\nu, \alpha(k'_\nu)}\delta_{i_\sigma(\nu),\beta(l'_\nu)})}^{r_\nu}\\
        ((-1)&\rho_{j'_{\tau(\nu)}i'_\nu}\sigma^\mu_{l_\nu i_{\sigma(\nu)}}O^i_{l'_\nu k_\nu}\delta_{k_\nu,\alpha(j'_\nu)}\delta_{l_\nu,\beta(l'_\nu)})^{1-r_\nu}\\
        :=\sum_{\substack{\sigma, \tau \in S_4\\
        \alpha, \beta}}& \text{Wg}^U_L(\sigma^{-1}\tau,d)\text{Wg}^U_R(\alpha^{-1}\beta,d)\Tr_{\sigma,\tau}^{\alpha,\beta}(\sigma^\mu,O^i,\rho)\\
        \leq \frac{C^2}{d^8}&\sum_{\substack{\sigma, \tau \in S_4\\ \alpha, \beta }}\Tr_{\sigma,\tau}^{\alpha,\beta}(\sigma^\mu,O^i,\rho).
     \end{split}
 \end{align}
Consequently, by substituting the Eq.~\eqref{C^4} and Eq.~\eqref{muC^4} in the Eq.~\eqref{Var C} we find the  variance $\text{Var}[\partial_\mu \mathcal{L}]$, is exponentially suppressed as a function of qubits number since 

\begin{align}
\begin{split}
    \text{Var}[\partial_\mu \mathcal{L}]\leq &\frac{C}{2^{6n}}\sum_{i,j=1}^{2^n}(q_i q_j)^{-1}[\sum_{\substack{\sigma', \tau' \in S_4\\ \alpha', \beta'}}\sum_{\substack{\sigma, \tau \in S_4\\ \alpha, \beta}}\Tr_{\sigma,\tau}^{\alpha,\beta}(\sigma^\mu,O^j,\rho)\\
    &\Tr_{\sigma, \tau}(\rho, O^i)\Tr_{\sigma',\tau'}^{\alpha',\beta'}(\sigma^\mu,O^i,\rho)\Tr_{\sigma', \tau'}(\rho, O^i)]^{\frac{1}{4}}
    \end{split}
\end{align}

\section{Bars and Strips ensemble }\label{appendixB}

The Bars and Stripes (BAS) \cite{mackay2003information}  ensemble is one of the standard benchmarks used to study the performance of  unsupervised generative models.
For a given  $(n,m)$, this data set, BAS$(n,m)$, can be constructed based on the two-dimensional grid with $n$ rows and $m$ columns where each plaquette  can be black (filled) or white (empty) with 
total number of possible configurations of $2^{n\cdot m}$. 
Each sample of BAS  belongs to a specific 
subset of these states which the grids are filled to make  exclusively $i$ filled columns ($0\leq i\leq m$) or exclusively $j$ rows ($0\leq j \leq n$), i.e. bars and stripes, respectively. Therefore, the total number of possible BAS samples is $N_{\text{BAS}(n,m)}\sum_{k=0}^{n} \binom{n}{k}+\sum_{p=0}^m \binom{m}{p}-2=2^{n}+2^{m}-2$.
The ratio of total to valid BAS patterns decreases exponentially as $2^{-\min(n,m)}$. 
To evaluate the performance of a quantum system to generate BAS states, a qBAS score was introuced \cite{benedetti2019generative}. It is defined as $\text{qBAS}=\frac{2pr}{p+r}$, where $r$ is the recall number, i.e. the ability to generate all patters of BAS$(n,m)$, and $p$ is the ability to retrieve states belongs to BAS$(n,m)$. In order to observe the whole spectrum of BAS$(n,m)$ patterns, we need to handful of measurements from the circuit. Since each pattern occurs with probability $1/N_{\text{BAS}(n,m)}$, we need to do $N_{M}=N_{\text{BAS}(n,m)} (1+\frac{1}{2}+\cdots \frac{1}{N_{\text{BAS}(n,m)}})\approx 2^{\max(m,n)}(\max(m,n)+\gamma)$ measurements where $\gamma$ is Euler-Macheroni constant.

\nocite{*}

\bibliography{main}

\begin{thebibliography}{73}%
\makeatletter
\providecommand \@ifxundefined [1]{%
 \@ifx{#1\undefined}
}%
\providecommand \@ifnum [1]{%
 \ifnum #1\expandafter \@firstoftwo
 \else \expandafter \@secondoftwo
 \fi
}%
\providecommand \@ifx [1]{%
 \ifx #1\expandafter \@firstoftwo
 \else \expandafter \@secondoftwo
 \fi
}%
\providecommand \natexlab [1]{#1}%
\providecommand \enquote  [1]{``#1''}%
\providecommand \bibnamefont  [1]{#1}%
\providecommand \bibfnamefont [1]{#1}%
\providecommand \citenamefont [1]{#1}%
\providecommand \href@noop [0]{\@secondoftwo}%
\providecommand \href [0]{\begingroup \@sanitize@url \@href}%
\providecommand \@href[1]{\@@startlink{#1}\@@href}%
\providecommand \@@href[1]{\endgroup#1\@@endlink}%
\providecommand \@sanitize@url [0]{\catcode `\\12\catcode `\$12\catcode
  `\&12\catcode `\#12\catcode `\^12\catcode `\_12\catcode `\%12\relax}%
\providecommand \@@startlink[1]{}%
\providecommand \@@endlink[0]{}%
\providecommand \url  [0]{\begingroup\@sanitize@url \@url }%
\providecommand \@url [1]{\endgroup\@href {#1}{\urlprefix }}%
\providecommand \urlprefix  [0]{URL }%
\providecommand \Eprint [0]{\href }%
\providecommand \doibase [0]{https://doi.org/}%
\providecommand \selectlanguage [0]{\@gobble}%
\providecommand \bibinfo  [0]{\@secondoftwo}%
\providecommand \bibfield  [0]{\@secondoftwo}%
\providecommand \translation [1]{[#1]}%
\providecommand \BibitemOpen [0]{}%
\providecommand \bibitemStop [0]{}%
\providecommand \bibitemNoStop [0]{.\EOS\space}%
\providecommand \EOS [0]{\spacefactor3000\relax}%
\providecommand \BibitemShut  [1]{\csname bibitem#1\endcsname}%
\let\auto@bib@innerbib\@empty
\bibitem [{\citenamefont {Cerezo}\ \emph
  {et~al.}(2021{\natexlab{a}})\citenamefont {Cerezo}, \citenamefont
  {Arrasmith}, \citenamefont {Babbush}, \citenamefont {Benjamin}, \citenamefont
  {Endo}, \citenamefont {Fujii}, \citenamefont {McClean}, \citenamefont
  {Mitarai}, \citenamefont {Yuan}, \citenamefont {Cincio} \emph
  {et~al.}}]{cerezo2021variational}%
  \BibitemOpen
  \bibfield  {author} {\bibinfo {author} {\bibfnamefont {M.}~\bibnamefont
  {Cerezo}}, \bibinfo {author} {\bibfnamefont {A.}~\bibnamefont {Arrasmith}},
  \bibinfo {author} {\bibfnamefont {R.}~\bibnamefont {Babbush}}, \bibinfo
  {author} {\bibfnamefont {S.~C.}\ \bibnamefont {Benjamin}}, \bibinfo {author}
  {\bibfnamefont {S.}~\bibnamefont {Endo}}, \bibinfo {author} {\bibfnamefont
  {K.}~\bibnamefont {Fujii}}, \bibinfo {author} {\bibfnamefont {J.~R.}\
  \bibnamefont {McClean}}, \bibinfo {author} {\bibfnamefont {K.}~\bibnamefont
  {Mitarai}}, \bibinfo {author} {\bibfnamefont {X.}~\bibnamefont {Yuan}},
  \bibinfo {author} {\bibfnamefont {L.}~\bibnamefont {Cincio}}, \emph
  {et~al.},\ }\bibfield  {title} {\bibinfo {title} {Variational quantum
  algorithms},\ }\href@noop {} {\bibfield  {journal} {\bibinfo  {journal}
  {Nature Reviews Physics}\ }\textbf {\bibinfo {volume} {3}},\ \bibinfo {pages}
  {625} (\bibinfo {year} {2021}{\natexlab{a}})}\BibitemShut {NoStop}%
\bibitem [{\citenamefont {McClean}\ \emph {et~al.}(2016)\citenamefont
  {McClean}, \citenamefont {Romero}, \citenamefont {Babbush},\ and\
  \citenamefont {Aspuru-Guzik}}]{mcclean2016theory}%
  \BibitemOpen
  \bibfield  {author} {\bibinfo {author} {\bibfnamefont {J.~R.}\ \bibnamefont
  {McClean}}, \bibinfo {author} {\bibfnamefont {J.}~\bibnamefont {Romero}},
  \bibinfo {author} {\bibfnamefont {R.}~\bibnamefont {Babbush}},\ and\ \bibinfo
  {author} {\bibfnamefont {A.}~\bibnamefont {Aspuru-Guzik}},\ }\bibfield
  {title} {\bibinfo {title} {The theory of variational hybrid quantum-classical
  algorithms},\ }\href@noop {} {\bibfield  {journal} {\bibinfo  {journal} {New
  Journal of Physics}\ }\textbf {\bibinfo {volume} {18}},\ \bibinfo {pages}
  {023023} (\bibinfo {year} {2016})}\BibitemShut {NoStop}%
\bibitem [{\citenamefont {Peruzzo}\ \emph {et~al.}(2014)\citenamefont
  {Peruzzo}, \citenamefont {McClean}, \citenamefont {Shadbolt}, \citenamefont
  {Yung}, \citenamefont {Zhou}, \citenamefont {Love}, \citenamefont
  {Aspuru-Guzik},\ and\ \citenamefont {O’brien}}]{peruzzo2014variational}%
  \BibitemOpen
  \bibfield  {author} {\bibinfo {author} {\bibfnamefont {A.}~\bibnamefont
  {Peruzzo}}, \bibinfo {author} {\bibfnamefont {J.}~\bibnamefont {McClean}},
  \bibinfo {author} {\bibfnamefont {P.}~\bibnamefont {Shadbolt}}, \bibinfo
  {author} {\bibfnamefont {M.-H.}\ \bibnamefont {Yung}}, \bibinfo {author}
  {\bibfnamefont {X.-Q.}\ \bibnamefont {Zhou}}, \bibinfo {author}
  {\bibfnamefont {P.~J.}\ \bibnamefont {Love}}, \bibinfo {author}
  {\bibfnamefont {A.}~\bibnamefont {Aspuru-Guzik}},\ and\ \bibinfo {author}
  {\bibfnamefont {J.~L.}\ \bibnamefont {O’brien}},\ }\bibfield  {title}
  {\bibinfo {title} {A variational eigenvalue solver on a photonic quantum
  processor},\ }\href@noop {} {\bibfield  {journal} {\bibinfo  {journal}
  {Nature communications}\ }\textbf {\bibinfo {volume} {5}},\ \bibinfo {pages}
  {1} (\bibinfo {year} {2014})}\BibitemShut {NoStop}%
\bibitem [{\citenamefont {Farhi}\ \emph {et~al.}(2014)\citenamefont {Farhi},
  \citenamefont {Goldstone},\ and\ \citenamefont {Gutmann}}]{farhi2014quantum}%
  \BibitemOpen
  \bibfield  {author} {\bibinfo {author} {\bibfnamefont {E.}~\bibnamefont
  {Farhi}}, \bibinfo {author} {\bibfnamefont {J.}~\bibnamefont {Goldstone}},\
  and\ \bibinfo {author} {\bibfnamefont {S.}~\bibnamefont {Gutmann}},\
  }\bibfield  {title} {\bibinfo {title} {A quantum approximate optimization
  algorithm},\ }\href@noop {} {\bibfield  {journal} {\bibinfo  {journal} {arXiv
  preprint arXiv:1411.4028}\ } (\bibinfo {year} {2014})}\BibitemShut {NoStop}%
\bibitem [{\citenamefont {Schuld}\ and\ \citenamefont
  {Killoran}(2019)}]{schuld2019quantum}%
  \BibitemOpen
  \bibfield  {author} {\bibinfo {author} {\bibfnamefont {M.}~\bibnamefont
  {Schuld}}\ and\ \bibinfo {author} {\bibfnamefont {N.}~\bibnamefont
  {Killoran}},\ }\bibfield  {title} {\bibinfo {title} {Quantum machine learning
  in feature hilbert spaces},\ }\href@noop {} {\bibfield  {journal} {\bibinfo
  {journal} {Physical review letters}\ }\textbf {\bibinfo {volume} {122}},\
  \bibinfo {pages} {040504} (\bibinfo {year} {2019})}\BibitemShut {NoStop}%
\bibitem [{\citenamefont {Schuld}\ \emph {et~al.}(2014)\citenamefont {Schuld},
  \citenamefont {Sinayskiy},\ and\ \citenamefont
  {Petruccione}}]{schuld2014quest}%
  \BibitemOpen
  \bibfield  {author} {\bibinfo {author} {\bibfnamefont {M.}~\bibnamefont
  {Schuld}}, \bibinfo {author} {\bibfnamefont {I.}~\bibnamefont {Sinayskiy}},\
  and\ \bibinfo {author} {\bibfnamefont {F.}~\bibnamefont {Petruccione}},\
  }\bibfield  {title} {\bibinfo {title} {The quest for a quantum neural
  network},\ }\href@noop {} {\bibfield  {journal} {\bibinfo  {journal} {Quantum
  Information Processing}\ }\textbf {\bibinfo {volume} {13}},\ \bibinfo {pages}
  {2567} (\bibinfo {year} {2014})}\BibitemShut {NoStop}%
\bibitem [{\citenamefont {Wiebe}\ \emph {et~al.}(2015)\citenamefont {Wiebe},
  \citenamefont {Kapoor},\ and\ \citenamefont {Svore}}]{wiebe2015quantum}%
  \BibitemOpen
  \bibfield  {author} {\bibinfo {author} {\bibfnamefont {N.}~\bibnamefont
  {Wiebe}}, \bibinfo {author} {\bibfnamefont {A.}~\bibnamefont {Kapoor}},\ and\
  \bibinfo {author} {\bibfnamefont {K.~M.}\ \bibnamefont {Svore}},\ }\bibfield
  {title} {\bibinfo {title} {Quantum nearest-neighbor algorithms for machine
  learning},\ }\href@noop {} {\bibfield  {journal} {\bibinfo  {journal}
  {Quantum information and computation}\ }\textbf {\bibinfo {volume} {15}},\
  \bibinfo {pages} {318} (\bibinfo {year} {2015})}\BibitemShut {NoStop}%
\bibitem [{\citenamefont {Kandala}\ \emph {et~al.}(2017)\citenamefont
  {Kandala}, \citenamefont {Mezzacapo}, \citenamefont {Temme}, \citenamefont
  {Takita}, \citenamefont {Brink}, \citenamefont {Chow},\ and\ \citenamefont
  {Gambetta}}]{kandala2017hardware}%
  \BibitemOpen
  \bibfield  {author} {\bibinfo {author} {\bibfnamefont {A.}~\bibnamefont
  {Kandala}}, \bibinfo {author} {\bibfnamefont {A.}~\bibnamefont {Mezzacapo}},
  \bibinfo {author} {\bibfnamefont {K.}~\bibnamefont {Temme}}, \bibinfo
  {author} {\bibfnamefont {M.}~\bibnamefont {Takita}}, \bibinfo {author}
  {\bibfnamefont {M.}~\bibnamefont {Brink}}, \bibinfo {author} {\bibfnamefont
  {J.~M.}\ \bibnamefont {Chow}},\ and\ \bibinfo {author} {\bibfnamefont
  {J.~M.}\ \bibnamefont {Gambetta}},\ }\bibfield  {title} {\bibinfo {title}
  {Hardware-efficient variational quantum eigensolver for small molecules and
  quantum magnets},\ }\href@noop {} {\bibfield  {journal} {\bibinfo  {journal}
  {Nature}\ }\textbf {\bibinfo {volume} {549}},\ \bibinfo {pages} {242}
  (\bibinfo {year} {2017})}\BibitemShut {NoStop}%
\bibitem [{\citenamefont {Dumitrescu}\ \emph {et~al.}(2018)\citenamefont
  {Dumitrescu}, \citenamefont {McCaskey}, \citenamefont {Hagen}, \citenamefont
  {Jansen}, \citenamefont {Morris}, \citenamefont {Papenbrock}, \citenamefont
  {Pooser}, \citenamefont {Dean},\ and\ \citenamefont
  {Lougovski}}]{dumitrescu2018cloud}%
  \BibitemOpen
  \bibfield  {author} {\bibinfo {author} {\bibfnamefont {E.~F.}\ \bibnamefont
  {Dumitrescu}}, \bibinfo {author} {\bibfnamefont {A.~J.}\ \bibnamefont
  {McCaskey}}, \bibinfo {author} {\bibfnamefont {G.}~\bibnamefont {Hagen}},
  \bibinfo {author} {\bibfnamefont {G.~R.}\ \bibnamefont {Jansen}}, \bibinfo
  {author} {\bibfnamefont {T.~D.}\ \bibnamefont {Morris}}, \bibinfo {author}
  {\bibfnamefont {T.}~\bibnamefont {Papenbrock}}, \bibinfo {author}
  {\bibfnamefont {R.~C.}\ \bibnamefont {Pooser}}, \bibinfo {author}
  {\bibfnamefont {D.~J.}\ \bibnamefont {Dean}},\ and\ \bibinfo {author}
  {\bibfnamefont {P.}~\bibnamefont {Lougovski}},\ }\bibfield  {title} {\bibinfo
  {title} {Cloud quantum computing of an atomic nucleus},\ }\href@noop {}
  {\bibfield  {journal} {\bibinfo  {journal} {Physical review letters}\
  }\textbf {\bibinfo {volume} {120}},\ \bibinfo {pages} {210501} (\bibinfo
  {year} {2018})}\BibitemShut {NoStop}%
\bibitem [{\citenamefont {Hempel}\ \emph {et~al.}(2018)\citenamefont {Hempel},
  \citenamefont {Maier}, \citenamefont {Romero}, \citenamefont {McClean},
  \citenamefont {Monz}, \citenamefont {Shen}, \citenamefont {Jurcevic},
  \citenamefont {Lanyon}, \citenamefont {Love}, \citenamefont {Babbush} \emph
  {et~al.}}]{hempel2018quantum}%
  \BibitemOpen
  \bibfield  {author} {\bibinfo {author} {\bibfnamefont {C.}~\bibnamefont
  {Hempel}}, \bibinfo {author} {\bibfnamefont {C.}~\bibnamefont {Maier}},
  \bibinfo {author} {\bibfnamefont {J.}~\bibnamefont {Romero}}, \bibinfo
  {author} {\bibfnamefont {J.}~\bibnamefont {McClean}}, \bibinfo {author}
  {\bibfnamefont {T.}~\bibnamefont {Monz}}, \bibinfo {author} {\bibfnamefont
  {H.}~\bibnamefont {Shen}}, \bibinfo {author} {\bibfnamefont {P.}~\bibnamefont
  {Jurcevic}}, \bibinfo {author} {\bibfnamefont {B.~P.}\ \bibnamefont
  {Lanyon}}, \bibinfo {author} {\bibfnamefont {P.}~\bibnamefont {Love}},
  \bibinfo {author} {\bibfnamefont {R.}~\bibnamefont {Babbush}}, \emph
  {et~al.},\ }\bibfield  {title} {\bibinfo {title} {Quantum chemistry
  calculations on a trapped-ion quantum simulator},\ }\href@noop {} {\bibfield
  {journal} {\bibinfo  {journal} {Physical Review X}\ }\textbf {\bibinfo
  {volume} {8}},\ \bibinfo {pages} {031022} (\bibinfo {year}
  {2018})}\BibitemShut {NoStop}%
\bibitem [{\citenamefont {O’Malley}\ \emph {et~al.}(2016)\citenamefont
  {O’Malley}, \citenamefont {Babbush}, \citenamefont {Kivlichan},
  \citenamefont {Romero}, \citenamefont {McClean}, \citenamefont {Barends},
  \citenamefont {Kelly}, \citenamefont {Roushan}, \citenamefont {Tranter},
  \citenamefont {Ding} \emph {et~al.}}]{o2016scalable}%
  \BibitemOpen
  \bibfield  {author} {\bibinfo {author} {\bibfnamefont {P.~J.}\ \bibnamefont
  {O’Malley}}, \bibinfo {author} {\bibfnamefont {R.}~\bibnamefont {Babbush}},
  \bibinfo {author} {\bibfnamefont {I.~D.}\ \bibnamefont {Kivlichan}}, \bibinfo
  {author} {\bibfnamefont {J.}~\bibnamefont {Romero}}, \bibinfo {author}
  {\bibfnamefont {J.~R.}\ \bibnamefont {McClean}}, \bibinfo {author}
  {\bibfnamefont {R.}~\bibnamefont {Barends}}, \bibinfo {author} {\bibfnamefont
  {J.}~\bibnamefont {Kelly}}, \bibinfo {author} {\bibfnamefont
  {P.}~\bibnamefont {Roushan}}, \bibinfo {author} {\bibfnamefont
  {A.}~\bibnamefont {Tranter}}, \bibinfo {author} {\bibfnamefont
  {N.}~\bibnamefont {Ding}}, \emph {et~al.},\ }\bibfield  {title} {\bibinfo
  {title} {Scalable quantum simulation of molecular energies},\ }\href@noop {}
  {\bibfield  {journal} {\bibinfo  {journal} {Physical Review X}\ }\textbf
  {\bibinfo {volume} {6}},\ \bibinfo {pages} {031007} (\bibinfo {year}
  {2016})}\BibitemShut {NoStop}%
\bibitem [{\citenamefont {Kokail}\ \emph {et~al.}(2019)\citenamefont {Kokail},
  \citenamefont {Maier}, \citenamefont {van Bijnen}, \citenamefont {Brydges},
  \citenamefont {Joshi}, \citenamefont {Jurcevic}, \citenamefont {Muschik},
  \citenamefont {Silvi}, \citenamefont {Blatt}, \citenamefont {Roos} \emph
  {et~al.}}]{kokail2019self}%
  \BibitemOpen
  \bibfield  {author} {\bibinfo {author} {\bibfnamefont {C.}~\bibnamefont
  {Kokail}}, \bibinfo {author} {\bibfnamefont {C.}~\bibnamefont {Maier}},
  \bibinfo {author} {\bibfnamefont {R.}~\bibnamefont {van Bijnen}}, \bibinfo
  {author} {\bibfnamefont {T.}~\bibnamefont {Brydges}}, \bibinfo {author}
  {\bibfnamefont {M.~K.}\ \bibnamefont {Joshi}}, \bibinfo {author}
  {\bibfnamefont {P.}~\bibnamefont {Jurcevic}}, \bibinfo {author}
  {\bibfnamefont {C.~A.}\ \bibnamefont {Muschik}}, \bibinfo {author}
  {\bibfnamefont {P.}~\bibnamefont {Silvi}}, \bibinfo {author} {\bibfnamefont
  {R.}~\bibnamefont {Blatt}}, \bibinfo {author} {\bibfnamefont {C.~F.}\
  \bibnamefont {Roos}}, \emph {et~al.},\ }\bibfield  {title} {\bibinfo {title}
  {Self-verifying variational quantum simulation of lattice models},\
  }\href@noop {} {\bibfield  {journal} {\bibinfo  {journal} {Nature}\ }\textbf
  {\bibinfo {volume} {569}},\ \bibinfo {pages} {355} (\bibinfo {year}
  {2019})}\BibitemShut {NoStop}%
\bibitem [{\citenamefont {Otterbach}\ \emph {et~al.}(2017)\citenamefont
  {Otterbach}, \citenamefont {Manenti}, \citenamefont {Alidoust}, \citenamefont
  {Bestwick}, \citenamefont {Block}, \citenamefont {Bloom}, \citenamefont
  {Caldwell}, \citenamefont {Didier}, \citenamefont {Fried}, \citenamefont
  {Hong} \emph {et~al.}}]{otterbach2017unsupervised}%
  \BibitemOpen
  \bibfield  {author} {\bibinfo {author} {\bibfnamefont {J.}~\bibnamefont
  {Otterbach}}, \bibinfo {author} {\bibfnamefont {R.}~\bibnamefont {Manenti}},
  \bibinfo {author} {\bibfnamefont {N.}~\bibnamefont {Alidoust}}, \bibinfo
  {author} {\bibfnamefont {A.}~\bibnamefont {Bestwick}}, \bibinfo {author}
  {\bibfnamefont {M.}~\bibnamefont {Block}}, \bibinfo {author} {\bibfnamefont
  {B.}~\bibnamefont {Bloom}}, \bibinfo {author} {\bibfnamefont
  {S.}~\bibnamefont {Caldwell}}, \bibinfo {author} {\bibfnamefont
  {N.}~\bibnamefont {Didier}}, \bibinfo {author} {\bibfnamefont {E.~S.}\
  \bibnamefont {Fried}}, \bibinfo {author} {\bibfnamefont {S.}~\bibnamefont
  {Hong}}, \emph {et~al.},\ }\bibfield  {title} {\bibinfo {title} {Unsupervised
  machine learning on a hybrid quantum computer},\ }\href@noop {} {\bibfield
  {journal} {\bibinfo  {journal} {arXiv preprint arXiv:1712.05771}\ } (\bibinfo
  {year} {2017})}\BibitemShut {NoStop}%
\bibitem [{\citenamefont {Zhu}\ \emph {et~al.}(2019)\citenamefont {Zhu},
  \citenamefont {Linke}, \citenamefont {Benedetti}, \citenamefont {Landsman},
  \citenamefont {Nguyen}, \citenamefont {Alderete}, \citenamefont
  {Perdomo-Ortiz}, \citenamefont {Korda}, \citenamefont {Garfoot},
  \citenamefont {Brecque} \emph {et~al.}}]{zhu2019training}%
  \BibitemOpen
  \bibfield  {author} {\bibinfo {author} {\bibfnamefont {D.}~\bibnamefont
  {Zhu}}, \bibinfo {author} {\bibfnamefont {N.~M.}\ \bibnamefont {Linke}},
  \bibinfo {author} {\bibfnamefont {M.}~\bibnamefont {Benedetti}}, \bibinfo
  {author} {\bibfnamefont {K.~A.}\ \bibnamefont {Landsman}}, \bibinfo {author}
  {\bibfnamefont {N.~H.}\ \bibnamefont {Nguyen}}, \bibinfo {author}
  {\bibfnamefont {C.~H.}\ \bibnamefont {Alderete}}, \bibinfo {author}
  {\bibfnamefont {A.}~\bibnamefont {Perdomo-Ortiz}}, \bibinfo {author}
  {\bibfnamefont {N.}~\bibnamefont {Korda}}, \bibinfo {author} {\bibfnamefont
  {A.}~\bibnamefont {Garfoot}}, \bibinfo {author} {\bibfnamefont
  {C.}~\bibnamefont {Brecque}}, \emph {et~al.},\ }\bibfield  {title} {\bibinfo
  {title} {Training of quantum circuits on a hybrid quantum computer},\
  }\href@noop {} {\bibfield  {journal} {\bibinfo  {journal} {Science advances}\
  }\textbf {\bibinfo {volume} {5}},\ \bibinfo {pages} {eaaw9918} (\bibinfo
  {year} {2019})}\BibitemShut {NoStop}%
\bibitem [{\citenamefont {Bittel}\ and\ \citenamefont
  {Kliesch}(2021)}]{bittel2021training}%
  \BibitemOpen
  \bibfield  {author} {\bibinfo {author} {\bibfnamefont {L.}~\bibnamefont
  {Bittel}}\ and\ \bibinfo {author} {\bibfnamefont {M.}~\bibnamefont
  {Kliesch}},\ }\bibfield  {title} {\bibinfo {title} {Training variational
  quantum algorithms is np-hard--even for logarithmically many qubits and free
  fermionic systems},\ }\href@noop {} {\bibfield  {journal} {\bibinfo
  {journal} {arXiv preprint arXiv:2101.07267}\ } (\bibinfo {year}
  {2021})}\BibitemShut {NoStop}%
\bibitem [{\citenamefont {Larocca}\ \emph {et~al.}(2021)\citenamefont
  {Larocca}, \citenamefont {Ju}, \citenamefont {Garc{\'\i}a-Mart{\'\i}n},
  \citenamefont {Coles},\ and\ \citenamefont {Cerezo}}]{larocca2021theory}%
  \BibitemOpen
  \bibfield  {author} {\bibinfo {author} {\bibfnamefont {M.}~\bibnamefont
  {Larocca}}, \bibinfo {author} {\bibfnamefont {N.}~\bibnamefont {Ju}},
  \bibinfo {author} {\bibfnamefont {D.}~\bibnamefont
  {Garc{\'\i}a-Mart{\'\i}n}}, \bibinfo {author} {\bibfnamefont {P.~J.}\
  \bibnamefont {Coles}},\ and\ \bibinfo {author} {\bibfnamefont
  {M.}~\bibnamefont {Cerezo}},\ }\bibfield  {title} {\bibinfo {title} {Theory
  of overparametrization in quantum neural networks},\ }\href@noop {}
  {\bibfield  {journal} {\bibinfo  {journal} {arXiv preprint arXiv:2109.11676}\
  } (\bibinfo {year} {2021})}\BibitemShut {NoStop}%
\bibitem [{\citenamefont {Cerezo}\ \emph
  {et~al.}(2021{\natexlab{b}})\citenamefont {Cerezo}, \citenamefont {Sone},
  \citenamefont {Volkoff}, \citenamefont {Cincio},\ and\ \citenamefont
  {Coles}}]{cerezo2021cost}%
  \BibitemOpen
  \bibfield  {author} {\bibinfo {author} {\bibfnamefont {M.}~\bibnamefont
  {Cerezo}}, \bibinfo {author} {\bibfnamefont {A.}~\bibnamefont {Sone}},
  \bibinfo {author} {\bibfnamefont {T.}~\bibnamefont {Volkoff}}, \bibinfo
  {author} {\bibfnamefont {L.}~\bibnamefont {Cincio}},\ and\ \bibinfo {author}
  {\bibfnamefont {P.~J.}\ \bibnamefont {Coles}},\ }\bibfield  {title} {\bibinfo
  {title} {Cost function dependent barren plateaus in shallow parametrized
  quantum circuits},\ }\href@noop {} {\bibfield  {journal} {\bibinfo  {journal}
  {Nature communications}\ }\textbf {\bibinfo {volume} {12}},\ \bibinfo {pages}
  {1} (\bibinfo {year} {2021}{\natexlab{b}})}\BibitemShut {NoStop}%
\bibitem [{\citenamefont {Holmes}\ \emph
  {et~al.}(2021{\natexlab{a}})\citenamefont {Holmes}, \citenamefont {Sharma},
  \citenamefont {Cerezo},\ and\ \citenamefont {Coles}}]{holmes2021connecting}%
  \BibitemOpen
  \bibfield  {author} {\bibinfo {author} {\bibfnamefont {Z.}~\bibnamefont
  {Holmes}}, \bibinfo {author} {\bibfnamefont {K.}~\bibnamefont {Sharma}},
  \bibinfo {author} {\bibfnamefont {M.}~\bibnamefont {Cerezo}},\ and\ \bibinfo
  {author} {\bibfnamefont {P.~J.}\ \bibnamefont {Coles}},\ }\bibfield  {title}
  {\bibinfo {title} {Connecting ansatz expressibility to gradient magnitudes
  and barren plateaus},\ }\href@noop {} {\bibfield  {journal} {\bibinfo
  {journal} {arXiv preprint arXiv:2101.02138}\ } (\bibinfo {year}
  {2021}{\natexlab{a}})}\BibitemShut {NoStop}%
\bibitem [{\citenamefont {Patti}\ \emph {et~al.}(2021)\citenamefont {Patti},
  \citenamefont {Najafi}, \citenamefont {Gao},\ and\ \citenamefont
  {Yelin}}]{patti2021entanglement}%
  \BibitemOpen
  \bibfield  {author} {\bibinfo {author} {\bibfnamefont {T.~L.}\ \bibnamefont
  {Patti}}, \bibinfo {author} {\bibfnamefont {K.}~\bibnamefont {Najafi}},
  \bibinfo {author} {\bibfnamefont {X.}~\bibnamefont {Gao}},\ and\ \bibinfo
  {author} {\bibfnamefont {S.~F.}\ \bibnamefont {Yelin}},\ }\bibfield  {title}
  {\bibinfo {title} {Entanglement devised barren plateau mitigation},\
  }\href@noop {} {\bibfield  {journal} {\bibinfo  {journal} {Physical Review
  Research}\ }\textbf {\bibinfo {volume} {3}},\ \bibinfo {pages} {033090}
  (\bibinfo {year} {2021})}\BibitemShut {NoStop}%
\bibitem [{\citenamefont {Marrero}\ \emph {et~al.}(2021)\citenamefont
  {Marrero}, \citenamefont {Kieferov{\'a}},\ and\ \citenamefont
  {Wiebe}}]{marrero2021entanglement}%
  \BibitemOpen
  \bibfield  {author} {\bibinfo {author} {\bibfnamefont {C.~O.}\ \bibnamefont
  {Marrero}}, \bibinfo {author} {\bibfnamefont {M.}~\bibnamefont
  {Kieferov{\'a}}},\ and\ \bibinfo {author} {\bibfnamefont {N.}~\bibnamefont
  {Wiebe}},\ }\bibfield  {title} {\bibinfo {title} {Entanglement-induced barren
  plateaus},\ }\href@noop {} {\bibfield  {journal} {\bibinfo  {journal} {PRX
  Quantum}\ }\textbf {\bibinfo {volume} {2}},\ \bibinfo {pages} {040316}
  (\bibinfo {year} {2021})}\BibitemShut {NoStop}%
\bibitem [{\citenamefont {Wang}\ \emph {et~al.}(2021)\citenamefont {Wang},
  \citenamefont {Fontana}, \citenamefont {Cerezo}, \citenamefont {Sharma},
  \citenamefont {Sone}, \citenamefont {Cincio},\ and\ \citenamefont
  {Coles}}]{wang2021noise}%
  \BibitemOpen
  \bibfield  {author} {\bibinfo {author} {\bibfnamefont {S.}~\bibnamefont
  {Wang}}, \bibinfo {author} {\bibfnamefont {E.}~\bibnamefont {Fontana}},
  \bibinfo {author} {\bibfnamefont {M.}~\bibnamefont {Cerezo}}, \bibinfo
  {author} {\bibfnamefont {K.}~\bibnamefont {Sharma}}, \bibinfo {author}
  {\bibfnamefont {A.}~\bibnamefont {Sone}}, \bibinfo {author} {\bibfnamefont
  {L.}~\bibnamefont {Cincio}},\ and\ \bibinfo {author} {\bibfnamefont {P.~J.}\
  \bibnamefont {Coles}},\ }\bibfield  {title} {\bibinfo {title} {Noise-induced
  barren plateaus in variational quantum algorithms},\ }\href@noop {}
  {\bibfield  {journal} {\bibinfo  {journal} {Nature Communications}\ }\textbf
  {\bibinfo {volume} {12}},\ \bibinfo {pages} {1} (\bibinfo {year}
  {2021})}\BibitemShut {NoStop}%
\bibitem [{\citenamefont {Du}\ \emph {et~al.}(2021)\citenamefont {Du},
  \citenamefont {Hsieh}, \citenamefont {Liu}, \citenamefont {You},\ and\
  \citenamefont {Tao}}]{du2021learnability}%
  \BibitemOpen
  \bibfield  {author} {\bibinfo {author} {\bibfnamefont {Y.}~\bibnamefont
  {Du}}, \bibinfo {author} {\bibfnamefont {M.-H.}\ \bibnamefont {Hsieh}},
  \bibinfo {author} {\bibfnamefont {T.}~\bibnamefont {Liu}}, \bibinfo {author}
  {\bibfnamefont {S.}~\bibnamefont {You}},\ and\ \bibinfo {author}
  {\bibfnamefont {D.}~\bibnamefont {Tao}},\ }\bibfield  {title} {\bibinfo
  {title} {Learnability of quantum neural networks},\ }\href@noop {} {\bibfield
   {journal} {\bibinfo  {journal} {PRX Quantum}\ }\textbf {\bibinfo {volume}
  {2}},\ \bibinfo {pages} {040337} (\bibinfo {year} {2021})}\BibitemShut
  {NoStop}%
\bibitem [{\citenamefont {Sharma}\ \emph {et~al.}(2020)\citenamefont {Sharma},
  \citenamefont {Cerezo}, \citenamefont {Cincio},\ and\ \citenamefont
  {Coles}}]{sharma2020trainability}%
  \BibitemOpen
  \bibfield  {author} {\bibinfo {author} {\bibfnamefont {K.}~\bibnamefont
  {Sharma}}, \bibinfo {author} {\bibfnamefont {M.}~\bibnamefont {Cerezo}},
  \bibinfo {author} {\bibfnamefont {L.}~\bibnamefont {Cincio}},\ and\ \bibinfo
  {author} {\bibfnamefont {P.~J.}\ \bibnamefont {Coles}},\ }\bibfield  {title}
  {\bibinfo {title} {Trainability of dissipative perceptron-based quantum
  neural networks},\ }\href@noop {} {\bibfield  {journal} {\bibinfo  {journal}
  {arXiv preprint arXiv:2005.12458}\ } (\bibinfo {year} {2020})}\BibitemShut
  {NoStop}%
\bibitem [{\citenamefont {McClean}\ \emph {et~al.}(2018)\citenamefont
  {McClean}, \citenamefont {Boixo}, \citenamefont {Smelyanskiy}, \citenamefont
  {Babbush},\ and\ \citenamefont {Neven}}]{mcclean2018barren}%
  \BibitemOpen
  \bibfield  {author} {\bibinfo {author} {\bibfnamefont {J.~R.}\ \bibnamefont
  {McClean}}, \bibinfo {author} {\bibfnamefont {S.}~\bibnamefont {Boixo}},
  \bibinfo {author} {\bibfnamefont {V.~N.}\ \bibnamefont {Smelyanskiy}},
  \bibinfo {author} {\bibfnamefont {R.}~\bibnamefont {Babbush}},\ and\ \bibinfo
  {author} {\bibfnamefont {H.}~\bibnamefont {Neven}},\ }\bibfield  {title}
  {\bibinfo {title} {Barren plateaus in quantum neural network training
  landscapes},\ }\href@noop {} {\bibfield  {journal} {\bibinfo  {journal}
  {Nature communications}\ }\textbf {\bibinfo {volume} {9}},\ \bibinfo {pages}
  {1} (\bibinfo {year} {2018})}\BibitemShut {NoStop}%
\bibitem [{\citenamefont {Guerreschi}\ and\ \citenamefont
  {Smelyanskiy}(2017)}]{guerreschi2017practical}%
  \BibitemOpen
  \bibfield  {author} {\bibinfo {author} {\bibfnamefont {G.~G.}\ \bibnamefont
  {Guerreschi}}\ and\ \bibinfo {author} {\bibfnamefont {M.}~\bibnamefont
  {Smelyanskiy}},\ }\bibfield  {title} {\bibinfo {title} {Practical
  optimization for hybrid quantum-classical algorithms},\ }\href@noop {}
  {\bibfield  {journal} {\bibinfo  {journal} {arXiv preprint arXiv:1701.01450}\
  } (\bibinfo {year} {2017})}\BibitemShut {NoStop}%
\bibitem [{\citenamefont {Schuld}\ \emph {et~al.}(2019)\citenamefont {Schuld},
  \citenamefont {Bergholm}, \citenamefont {Gogolin}, \citenamefont {Izaac},\
  and\ \citenamefont {Killoran}}]{schuld2019evaluating}%
  \BibitemOpen
  \bibfield  {author} {\bibinfo {author} {\bibfnamefont {M.}~\bibnamefont
  {Schuld}}, \bibinfo {author} {\bibfnamefont {V.}~\bibnamefont {Bergholm}},
  \bibinfo {author} {\bibfnamefont {C.}~\bibnamefont {Gogolin}}, \bibinfo
  {author} {\bibfnamefont {J.}~\bibnamefont {Izaac}},\ and\ \bibinfo {author}
  {\bibfnamefont {N.}~\bibnamefont {Killoran}},\ }\bibfield  {title} {\bibinfo
  {title} {Evaluating analytic gradients on quantum hardware},\ }\href@noop {}
  {\bibfield  {journal} {\bibinfo  {journal} {Physical Review A}\ }\textbf
  {\bibinfo {volume} {99}},\ \bibinfo {pages} {032331} (\bibinfo {year}
  {2019})}\BibitemShut {NoStop}%
\bibitem [{\citenamefont {Mari}\ \emph {et~al.}(2021)\citenamefont {Mari},
  \citenamefont {Bromley},\ and\ \citenamefont
  {Killoran}}]{mari2021estimating}%
  \BibitemOpen
  \bibfield  {author} {\bibinfo {author} {\bibfnamefont {A.}~\bibnamefont
  {Mari}}, \bibinfo {author} {\bibfnamefont {T.~R.}\ \bibnamefont {Bromley}},\
  and\ \bibinfo {author} {\bibfnamefont {N.}~\bibnamefont {Killoran}},\
  }\bibfield  {title} {\bibinfo {title} {Estimating the gradient and
  higher-order derivatives on quantum hardware},\ }\href@noop {} {\bibfield
  {journal} {\bibinfo  {journal} {Physical Review A}\ }\textbf {\bibinfo
  {volume} {103}},\ \bibinfo {pages} {012405} (\bibinfo {year}
  {2021})}\BibitemShut {NoStop}%
\bibitem [{\citenamefont {Cerezo}\ and\ \citenamefont
  {Coles}(2021)}]{cerezo2021higher}%
  \BibitemOpen
  \bibfield  {author} {\bibinfo {author} {\bibfnamefont {M.}~\bibnamefont
  {Cerezo}}\ and\ \bibinfo {author} {\bibfnamefont {P.~J.}\ \bibnamefont
  {Coles}},\ }\bibfield  {title} {\bibinfo {title} {Higher order derivatives of
  quantum neural networks with barren plateaus},\ }\href@noop {} {\bibfield
  {journal} {\bibinfo  {journal} {Quantum Science and Technology}\ }\textbf
  {\bibinfo {volume} {6}},\ \bibinfo {pages} {035006} (\bibinfo {year}
  {2021})}\BibitemShut {NoStop}%
\bibitem [{\citenamefont {Cerezo}\ and\ \citenamefont
  {Coles}(2020)}]{cerezo2020impact}%
  \BibitemOpen
  \bibfield  {author} {\bibinfo {author} {\bibfnamefont {M.}~\bibnamefont
  {Cerezo}}\ and\ \bibinfo {author} {\bibfnamefont {P.~J.}\ \bibnamefont
  {Coles}},\ }\bibfield  {title} {\bibinfo {title} {Impact of barren plateaus
  on the hessian and higher order derivatives},\ }\href@noop {} {\bibfield
  {journal} {\bibinfo  {journal} {arXiv preprint arXiv:2008.07454}\ } (\bibinfo
  {year} {2020})}\BibitemShut {NoStop}%
\bibitem [{\citenamefont {Sweke}\ \emph {et~al.}(2020)\citenamefont {Sweke},
  \citenamefont {Wilde}, \citenamefont {Meyer}, \citenamefont {Schuld},
  \citenamefont {F{\"a}hrmann}, \citenamefont {Meynard-Piganeau},\ and\
  \citenamefont {Eisert}}]{sweke2020stochastic}%
  \BibitemOpen
  \bibfield  {author} {\bibinfo {author} {\bibfnamefont {R.}~\bibnamefont
  {Sweke}}, \bibinfo {author} {\bibfnamefont {F.}~\bibnamefont {Wilde}},
  \bibinfo {author} {\bibfnamefont {J.}~\bibnamefont {Meyer}}, \bibinfo
  {author} {\bibfnamefont {M.}~\bibnamefont {Schuld}}, \bibinfo {author}
  {\bibfnamefont {P.~K.}\ \bibnamefont {F{\"a}hrmann}}, \bibinfo {author}
  {\bibfnamefont {B.}~\bibnamefont {Meynard-Piganeau}},\ and\ \bibinfo {author}
  {\bibfnamefont {J.}~\bibnamefont {Eisert}},\ }\bibfield  {title} {\bibinfo
  {title} {Stochastic gradient descent for hybrid quantum-classical
  optimization},\ }\href@noop {} {\bibfield  {journal} {\bibinfo  {journal}
  {Quantum}\ }\textbf {\bibinfo {volume} {4}},\ \bibinfo {pages} {314}
  (\bibinfo {year} {2020})}\BibitemShut {NoStop}%
\bibitem [{\citenamefont {Stokes}\ \emph {et~al.}(2020)\citenamefont {Stokes},
  \citenamefont {Izaac}, \citenamefont {Killoran},\ and\ \citenamefont
  {Carleo}}]{stokes2020quantum}%
  \BibitemOpen
  \bibfield  {author} {\bibinfo {author} {\bibfnamefont {J.}~\bibnamefont
  {Stokes}}, \bibinfo {author} {\bibfnamefont {J.}~\bibnamefont {Izaac}},
  \bibinfo {author} {\bibfnamefont {N.}~\bibnamefont {Killoran}},\ and\
  \bibinfo {author} {\bibfnamefont {G.}~\bibnamefont {Carleo}},\ }\bibfield
  {title} {\bibinfo {title} {Quantum natural gradient},\ }\href@noop {}
  {\bibfield  {journal} {\bibinfo  {journal} {Quantum}\ }\textbf {\bibinfo
  {volume} {4}},\ \bibinfo {pages} {269} (\bibinfo {year} {2020})}\BibitemShut
  {NoStop}%
\bibitem [{\citenamefont {K{\"u}bler}\ \emph {et~al.}(2020)\citenamefont
  {K{\"u}bler}, \citenamefont {Arrasmith}, \citenamefont {Cincio},\ and\
  \citenamefont {Coles}}]{kubler2020adaptive}%
  \BibitemOpen
  \bibfield  {author} {\bibinfo {author} {\bibfnamefont {J.~M.}\ \bibnamefont
  {K{\"u}bler}}, \bibinfo {author} {\bibfnamefont {A.}~\bibnamefont
  {Arrasmith}}, \bibinfo {author} {\bibfnamefont {L.}~\bibnamefont {Cincio}},\
  and\ \bibinfo {author} {\bibfnamefont {P.~J.}\ \bibnamefont {Coles}},\
  }\bibfield  {title} {\bibinfo {title} {An adaptive optimizer for
  measurement-frugal variational algorithms},\ }\href@noop {} {\bibfield
  {journal} {\bibinfo  {journal} {Quantum}\ }\textbf {\bibinfo {volume} {4}},\
  \bibinfo {pages} {263} (\bibinfo {year} {2020})}\BibitemShut {NoStop}%
\bibitem [{\citenamefont {Wilson}\ \emph {et~al.}(2021)\citenamefont {Wilson},
  \citenamefont {Stromswold}, \citenamefont {Wudarski}, \citenamefont
  {Hadfield}, \citenamefont {Tubman},\ and\ \citenamefont
  {Rieffel}}]{wilson2021optimizing}%
  \BibitemOpen
  \bibfield  {author} {\bibinfo {author} {\bibfnamefont {M.}~\bibnamefont
  {Wilson}}, \bibinfo {author} {\bibfnamefont {R.}~\bibnamefont {Stromswold}},
  \bibinfo {author} {\bibfnamefont {F.}~\bibnamefont {Wudarski}}, \bibinfo
  {author} {\bibfnamefont {S.}~\bibnamefont {Hadfield}}, \bibinfo {author}
  {\bibfnamefont {N.~M.}\ \bibnamefont {Tubman}},\ and\ \bibinfo {author}
  {\bibfnamefont {E.~G.}\ \bibnamefont {Rieffel}},\ }\bibfield  {title}
  {\bibinfo {title} {Optimizing quantum heuristics with meta-learning},\
  }\href@noop {} {\bibfield  {journal} {\bibinfo  {journal} {Quantum Machine
  Intelligence}\ }\textbf {\bibinfo {volume} {3}},\ \bibinfo {pages} {1}
  (\bibinfo {year} {2021})}\BibitemShut {NoStop}%
\bibitem [{\citenamefont {Spall}\ \emph {et~al.}(1992)\citenamefont {Spall}
  \emph {et~al.}}]{spall1992multivariate}%
  \BibitemOpen
  \bibfield  {author} {\bibinfo {author} {\bibfnamefont {J.~C.}\ \bibnamefont
  {Spall}} \emph {et~al.},\ }\bibfield  {title} {\bibinfo {title} {Multivariate
  stochastic approximation using a simultaneous perturbation gradient
  approximation},\ }\href@noop {} {\bibfield  {journal} {\bibinfo  {journal}
  {IEEE transactions on automatic control}\ }\textbf {\bibinfo {volume} {37}},\
  \bibinfo {pages} {332} (\bibinfo {year} {1992})}\BibitemShut {NoStop}%
\bibitem [{\citenamefont {Powell}(1994)}]{powell1994direct}%
  \BibitemOpen
  \bibfield  {author} {\bibinfo {author} {\bibfnamefont {M.~J.}\ \bibnamefont
  {Powell}},\ }\bibfield  {title} {\bibinfo {title} {A direct search
  optimization method that models the objective and constraint functions by
  linear interpolation},\ }in\ \href@noop {} {\emph {\bibinfo {booktitle}
  {Advances in optimization and numerical analysis}}}\ (\bibinfo  {publisher}
  {Springer},\ \bibinfo {year} {1994})\ pp.\ \bibinfo {pages}
  {51--67}\BibitemShut {NoStop}%
\bibitem [{\citenamefont {Powell}(1964)}]{powell1964efficient}%
  \BibitemOpen
  \bibfield  {author} {\bibinfo {author} {\bibfnamefont {M.~J.}\ \bibnamefont
  {Powell}},\ }\bibfield  {title} {\bibinfo {title} {An efficient method for
  finding the minimum of a function of several variables without calculating
  derivatives},\ }\href@noop {} {\bibfield  {journal} {\bibinfo  {journal} {The
  computer journal}\ }\textbf {\bibinfo {volume} {7}},\ \bibinfo {pages} {155}
  (\bibinfo {year} {1964})}\BibitemShut {NoStop}%
\bibitem [{\citenamefont {Arrasmith}\ \emph {et~al.}(2021)\citenamefont
  {Arrasmith}, \citenamefont {Cerezo}, \citenamefont {Czarnik}, \citenamefont
  {Cincio},\ and\ \citenamefont {Coles}}]{arrasmith2021effect}%
  \BibitemOpen
  \bibfield  {author} {\bibinfo {author} {\bibfnamefont {A.}~\bibnamefont
  {Arrasmith}}, \bibinfo {author} {\bibfnamefont {M.}~\bibnamefont {Cerezo}},
  \bibinfo {author} {\bibfnamefont {P.}~\bibnamefont {Czarnik}}, \bibinfo
  {author} {\bibfnamefont {L.}~\bibnamefont {Cincio}},\ and\ \bibinfo {author}
  {\bibfnamefont {P.~J.}\ \bibnamefont {Coles}},\ }\bibfield  {title} {\bibinfo
  {title} {Effect of barren plateaus on gradient-free optimization},\
  }\href@noop {} {\bibfield  {journal} {\bibinfo  {journal} {Quantum}\ }\textbf
  {\bibinfo {volume} {5}},\ \bibinfo {pages} {558} (\bibinfo {year}
  {2021})}\BibitemShut {NoStop}%
\bibitem [{\citenamefont {Bonet-Monroig}\ \emph {et~al.}(2021)\citenamefont
  {Bonet-Monroig}, \citenamefont {Wang}, \citenamefont {Vermetten},
  \citenamefont {Senjean}, \citenamefont {Moussa}, \citenamefont {B{\"a}ck},
  \citenamefont {Dunjko},\ and\ \citenamefont
  {O'Brien}}]{bonet2021performance}%
  \BibitemOpen
  \bibfield  {author} {\bibinfo {author} {\bibfnamefont {X.}~\bibnamefont
  {Bonet-Monroig}}, \bibinfo {author} {\bibfnamefont {H.}~\bibnamefont {Wang}},
  \bibinfo {author} {\bibfnamefont {D.}~\bibnamefont {Vermetten}}, \bibinfo
  {author} {\bibfnamefont {B.}~\bibnamefont {Senjean}}, \bibinfo {author}
  {\bibfnamefont {C.}~\bibnamefont {Moussa}}, \bibinfo {author} {\bibfnamefont
  {T.}~\bibnamefont {B{\"a}ck}}, \bibinfo {author} {\bibfnamefont
  {V.}~\bibnamefont {Dunjko}},\ and\ \bibinfo {author} {\bibfnamefont {T.~E.}\
  \bibnamefont {O'Brien}},\ }\bibfield  {title} {\bibinfo {title} {Performance
  comparison of optimization methods on variational quantum algorithms},\
  }\href@noop {} {\bibfield  {journal} {\bibinfo  {journal} {arXiv preprint
  arXiv:2111.13454}\ } (\bibinfo {year} {2021})}\BibitemShut {NoStop}%
\bibitem [{\citenamefont {Verdon}\ \emph {et~al.}(2019)\citenamefont {Verdon},
  \citenamefont {Broughton}, \citenamefont {McClean}, \citenamefont {Sung},
  \citenamefont {Babbush}, \citenamefont {Jiang}, \citenamefont {Neven},\ and\
  \citenamefont {Mohseni}}]{verdon2019learning}%
  \BibitemOpen
  \bibfield  {author} {\bibinfo {author} {\bibfnamefont {G.}~\bibnamefont
  {Verdon}}, \bibinfo {author} {\bibfnamefont {M.}~\bibnamefont {Broughton}},
  \bibinfo {author} {\bibfnamefont {J.~R.}\ \bibnamefont {McClean}}, \bibinfo
  {author} {\bibfnamefont {K.~J.}\ \bibnamefont {Sung}}, \bibinfo {author}
  {\bibfnamefont {R.}~\bibnamefont {Babbush}}, \bibinfo {author} {\bibfnamefont
  {Z.}~\bibnamefont {Jiang}}, \bibinfo {author} {\bibfnamefont
  {H.}~\bibnamefont {Neven}},\ and\ \bibinfo {author} {\bibfnamefont
  {M.}~\bibnamefont {Mohseni}},\ }\bibfield  {title} {\bibinfo {title}
  {Learning to learn with quantum neural networks via classical neural
  networks},\ }\href@noop {} {\bibfield  {journal} {\bibinfo  {journal} {arXiv
  preprint arXiv:1907.05415}\ } (\bibinfo {year} {2019})}\BibitemShut {NoStop}%
\bibitem [{\citenamefont {Holmes}\ \emph
  {et~al.}(2021{\natexlab{b}})\citenamefont {Holmes}, \citenamefont
  {Arrasmith}, \citenamefont {Yan}, \citenamefont {Coles}, \citenamefont
  {Albrecht},\ and\ \citenamefont {Sornborger}}]{holmes2021barren}%
  \BibitemOpen
  \bibfield  {author} {\bibinfo {author} {\bibfnamefont {Z.}~\bibnamefont
  {Holmes}}, \bibinfo {author} {\bibfnamefont {A.}~\bibnamefont {Arrasmith}},
  \bibinfo {author} {\bibfnamefont {B.}~\bibnamefont {Yan}}, \bibinfo {author}
  {\bibfnamefont {P.~J.}\ \bibnamefont {Coles}}, \bibinfo {author}
  {\bibfnamefont {A.}~\bibnamefont {Albrecht}},\ and\ \bibinfo {author}
  {\bibfnamefont {A.~T.}\ \bibnamefont {Sornborger}},\ }\bibfield  {title}
  {\bibinfo {title} {Barren plateaus preclude learning scramblers},\
  }\href@noop {} {\bibfield  {journal} {\bibinfo  {journal} {Physical Review
  Letters}\ }\textbf {\bibinfo {volume} {126}},\ \bibinfo {pages} {190501}
  (\bibinfo {year} {2021}{\natexlab{b}})}\BibitemShut {NoStop}%
\bibitem [{\citenamefont {Zhang}\ \emph {et~al.}(2020)\citenamefont {Zhang},
  \citenamefont {Hsieh}, \citenamefont {Liu},\ and\ \citenamefont
  {Tao}}]{zhang2020toward}%
  \BibitemOpen
  \bibfield  {author} {\bibinfo {author} {\bibfnamefont {K.}~\bibnamefont
  {Zhang}}, \bibinfo {author} {\bibfnamefont {M.-H.}\ \bibnamefont {Hsieh}},
  \bibinfo {author} {\bibfnamefont {L.}~\bibnamefont {Liu}},\ and\ \bibinfo
  {author} {\bibfnamefont {D.}~\bibnamefont {Tao}},\ }\bibfield  {title}
  {\bibinfo {title} {Toward trainability of quantum neural networks},\
  }\href@noop {} {\bibfield  {journal} {\bibinfo  {journal} {arXiv preprint
  arXiv:2011.06258}\ } (\bibinfo {year} {2020})}\BibitemShut {NoStop}%
\bibitem [{\citenamefont {Uvarov}\ and\ \citenamefont
  {Biamonte}(2021)}]{uvarov2021barren}%
  \BibitemOpen
  \bibfield  {author} {\bibinfo {author} {\bibfnamefont {A.}~\bibnamefont
  {Uvarov}}\ and\ \bibinfo {author} {\bibfnamefont {J.~D.}\ \bibnamefont
  {Biamonte}},\ }\bibfield  {title} {\bibinfo {title} {On barren plateaus and
  cost function locality in variational quantum algorithms},\ }\href@noop {}
  {\bibfield  {journal} {\bibinfo  {journal} {Journal of Physics A:
  Mathematical and Theoretical}\ }\textbf {\bibinfo {volume} {54}},\ \bibinfo
  {pages} {245301} (\bibinfo {year} {2021})}\BibitemShut {NoStop}%
\bibitem [{\citenamefont {Grant}\ \emph {et~al.}(2019)\citenamefont {Grant},
  \citenamefont {Wossnig}, \citenamefont {Ostaszewski},\ and\ \citenamefont
  {Benedetti}}]{grant2019initialization}%
  \BibitemOpen
  \bibfield  {author} {\bibinfo {author} {\bibfnamefont {E.}~\bibnamefont
  {Grant}}, \bibinfo {author} {\bibfnamefont {L.}~\bibnamefont {Wossnig}},
  \bibinfo {author} {\bibfnamefont {M.}~\bibnamefont {Ostaszewski}},\ and\
  \bibinfo {author} {\bibfnamefont {M.}~\bibnamefont {Benedetti}},\ }\bibfield
  {title} {\bibinfo {title} {An initialization strategy for addressing barren
  plateaus in parametrized quantum circuits},\ }\href@noop {} {\bibfield
  {journal} {\bibinfo  {journal} {Quantum}\ }\textbf {\bibinfo {volume} {3}},\
  \bibinfo {pages} {214} (\bibinfo {year} {2019})}\BibitemShut {NoStop}%
\bibitem [{\citenamefont {Volkoff}\ and\ \citenamefont
  {Coles}(2021)}]{volkoff2021large}%
  \BibitemOpen
  \bibfield  {author} {\bibinfo {author} {\bibfnamefont {T.}~\bibnamefont
  {Volkoff}}\ and\ \bibinfo {author} {\bibfnamefont {P.~J.}\ \bibnamefont
  {Coles}},\ }\bibfield  {title} {\bibinfo {title} {Large gradients via
  correlation in random parameterized quantum circuits},\ }\href@noop {}
  {\bibfield  {journal} {\bibinfo  {journal} {Quantum Science and Technology}\
  }\textbf {\bibinfo {volume} {6}},\ \bibinfo {pages} {025008} (\bibinfo {year}
  {2021})}\BibitemShut {NoStop}%
\bibitem [{\citenamefont {McLeod}\ \emph {et~al.}(2018)\citenamefont {McLeod},
  \citenamefont {Roberts},\ and\ \citenamefont
  {Osborne}}]{mcleod2018optimization}%
  \BibitemOpen
  \bibfield  {author} {\bibinfo {author} {\bibfnamefont {M.}~\bibnamefont
  {McLeod}}, \bibinfo {author} {\bibfnamefont {S.}~\bibnamefont {Roberts}},\
  and\ \bibinfo {author} {\bibfnamefont {M.~A.}\ \bibnamefont {Osborne}},\
  }\bibfield  {title} {\bibinfo {title} {Optimization, fast and slow: optimally
  switching between local and bayesian optimization},\ }in\ \href@noop {}
  {\emph {\bibinfo {booktitle} {International Conference on Machine
  Learning}}}\ (\bibinfo {organization} {PMLR},\ \bibinfo {year} {2018})\ pp.\
  \bibinfo {pages} {3443--3452}\BibitemShut {NoStop}%
\bibitem [{\citenamefont {Benedetti}\ \emph {et~al.}(2019)\citenamefont
  {Benedetti}, \citenamefont {Garcia-Pintos}, \citenamefont {Perdomo},
  \citenamefont {Leyton-Ortega}, \citenamefont {Nam},\ and\ \citenamefont
  {Perdomo-Ortiz}}]{benedetti2019generative}%
  \BibitemOpen
  \bibfield  {author} {\bibinfo {author} {\bibfnamefont {M.}~\bibnamefont
  {Benedetti}}, \bibinfo {author} {\bibfnamefont {D.}~\bibnamefont
  {Garcia-Pintos}}, \bibinfo {author} {\bibfnamefont {O.}~\bibnamefont
  {Perdomo}}, \bibinfo {author} {\bibfnamefont {V.}~\bibnamefont
  {Leyton-Ortega}}, \bibinfo {author} {\bibfnamefont {Y.}~\bibnamefont {Nam}},\
  and\ \bibinfo {author} {\bibfnamefont {A.}~\bibnamefont {Perdomo-Ortiz}},\
  }\bibfield  {title} {\bibinfo {title} {A generative modeling approach for
  benchmarking and training shallow quantum circuits},\ }\href@noop {}
  {\bibfield  {journal} {\bibinfo  {journal} {npj Quantum Information}\
  }\textbf {\bibinfo {volume} {5}},\ \bibinfo {pages} {1} (\bibinfo {year}
  {2019})}\BibitemShut {NoStop}%
\bibitem [{\citenamefont {Frazier}(2018)}]{frazier2018tutorial}%
  \BibitemOpen
  \bibfield  {author} {\bibinfo {author} {\bibfnamefont {P.~I.}\ \bibnamefont
  {Frazier}},\ }\bibfield  {title} {\bibinfo {title} {A tutorial on bayesian
  optimization},\ }\href@noop {} {\bibfield  {journal} {\bibinfo  {journal}
  {arXiv preprint arXiv:1807.02811}\ } (\bibinfo {year} {2018})}\BibitemShut
  {NoStop}%
\bibitem [{\citenamefont {Shahriari}\ \emph {et~al.}(2015)\citenamefont
  {Shahriari}, \citenamefont {Swersky}, \citenamefont {Wang}, \citenamefont
  {Adams},\ and\ \citenamefont {De~Freitas}}]{shahriari2015taking}%
  \BibitemOpen
  \bibfield  {author} {\bibinfo {author} {\bibfnamefont {B.}~\bibnamefont
  {Shahriari}}, \bibinfo {author} {\bibfnamefont {K.}~\bibnamefont {Swersky}},
  \bibinfo {author} {\bibfnamefont {Z.}~\bibnamefont {Wang}}, \bibinfo {author}
  {\bibfnamefont {R.~P.}\ \bibnamefont {Adams}},\ and\ \bibinfo {author}
  {\bibfnamefont {N.}~\bibnamefont {De~Freitas}},\ }\bibfield  {title}
  {\bibinfo {title} {Taking the human out of the loop: A review of bayesian
  optimization},\ }\href@noop {} {\bibfield  {journal} {\bibinfo  {journal}
  {Proceedings of the IEEE}\ }\textbf {\bibinfo {volume} {104}},\ \bibinfo
  {pages} {148} (\bibinfo {year} {2015})}\BibitemShut {NoStop}%
\bibitem [{\citenamefont {Liu}\ \emph {et~al.}(2021)\citenamefont {Liu},
  \citenamefont {Tacchino}, \citenamefont {Glick}, \citenamefont {Jiang},\ and\
  \citenamefont {Mezzacapo}}]{liu2021representation}%
  \BibitemOpen
  \bibfield  {author} {\bibinfo {author} {\bibfnamefont {J.}~\bibnamefont
  {Liu}}, \bibinfo {author} {\bibfnamefont {F.}~\bibnamefont {Tacchino}},
  \bibinfo {author} {\bibfnamefont {J.~R.}\ \bibnamefont {Glick}}, \bibinfo
  {author} {\bibfnamefont {L.}~\bibnamefont {Jiang}},\ and\ \bibinfo {author}
  {\bibfnamefont {A.}~\bibnamefont {Mezzacapo}},\ }\bibfield  {title} {\bibinfo
  {title} {Representation learning via quantum neural tangent kernels},\
  }\href@noop {} {\bibfield  {journal} {\bibinfo  {journal} {arXiv preprint
  arXiv:2111.04225}\ } (\bibinfo {year} {2021})}\BibitemShut {NoStop}%
\bibitem [{\citenamefont {Barton}\ and\ \citenamefont
  {Ivey~Jr}(1991)}]{barton1991modifications}%
  \BibitemOpen
  \bibfield  {author} {\bibinfo {author} {\bibfnamefont {R.~R.}\ \bibnamefont
  {Barton}}\ and\ \bibinfo {author} {\bibfnamefont {J.~S.}\ \bibnamefont
  {Ivey~Jr}},\ }\href@noop {} {\emph {\bibinfo {title} {Modifications of the
  Nelder-Mead simplex method for stochastic simulation response
  optimization}}},\ \bibinfo {type} {Tech. Rep.}\ (\bibinfo  {institution}
  {Institute of Electrical and Electronics Engineers (IEEE)},\ \bibinfo {year}
  {1991})\BibitemShut {NoStop}%
\bibitem [{Note1()}]{Note1}%
  \BibitemOpen
  \bibinfo {note} {\protect \texttt {Fast and Slow Algorithm} codebase:
  \protect \url
  {https://github.com/frustea/Quantum-Fast-and-Slow}.}\BibitemShut {Stop}%
\bibitem [{\citenamefont {MacKay}\ and\ \citenamefont
  {Mac~Kay}(2003)}]{mackay2003information}%
  \BibitemOpen
  \bibfield  {author} {\bibinfo {author} {\bibfnamefont {D.~J.}\ \bibnamefont
  {MacKay}}\ and\ \bibinfo {author} {\bibfnamefont {D.~J.}\ \bibnamefont
  {Mac~Kay}},\ }\href@noop {} {\emph {\bibinfo {title} {Information theory,
  inference and learning algorithms}}}\ (\bibinfo  {publisher} {Cambridge
  university press},\ \bibinfo {year} {2003})\BibitemShut {NoStop}%
\bibitem [{\citenamefont {Kullback}\ and\ \citenamefont
  {Leibler}(1951)}]{kullback1951information}%
  \BibitemOpen
  \bibfield  {author} {\bibinfo {author} {\bibfnamefont {S.}~\bibnamefont
  {Kullback}}\ and\ \bibinfo {author} {\bibfnamefont {R.~A.}\ \bibnamefont
  {Leibler}},\ }\bibfield  {title} {\bibinfo {title} {On information and
  sufficiency},\ }\href@noop {} {\bibfield  {journal} {\bibinfo  {journal} {The
  annals of mathematical statistics}\ }\textbf {\bibinfo {volume} {22}},\
  \bibinfo {pages} {79} (\bibinfo {year} {1951})}\BibitemShut {NoStop}%
\bibitem [{\citenamefont {Shalev-Shwartz}\ and\ \citenamefont
  {Ben-David}(2014)}]{shalev2014understanding}%
  \BibitemOpen
  \bibfield  {author} {\bibinfo {author} {\bibfnamefont {S.}~\bibnamefont
  {Shalev-Shwartz}}\ and\ \bibinfo {author} {\bibfnamefont {S.}~\bibnamefont
  {Ben-David}},\ }\href@noop {} {\emph {\bibinfo {title} {Understanding machine
  learning: From theory to algorithms}}}\ (\bibinfo  {publisher} {Cambridge
  university press},\ \bibinfo {year} {2014})\BibitemShut {NoStop}%
\bibitem [{\citenamefont {Harrow}\ and\ \citenamefont
  {Napp}(2021)}]{harrow2021low}%
  \BibitemOpen
  \bibfield  {author} {\bibinfo {author} {\bibfnamefont {A.~W.}\ \bibnamefont
  {Harrow}}\ and\ \bibinfo {author} {\bibfnamefont {J.~C.}\ \bibnamefont
  {Napp}},\ }\bibfield  {title} {\bibinfo {title} {Low-depth gradient
  measurements can improve convergence in variational hybrid quantum-classical
  algorithms},\ }\href@noop {} {\bibfield  {journal} {\bibinfo  {journal}
  {Physical Review Letters}\ }\textbf {\bibinfo {volume} {126}},\ \bibinfo
  {pages} {140502} (\bibinfo {year} {2021})}\BibitemShut {NoStop}%
\bibitem [{\citenamefont {Nelder}\ and\ \citenamefont
  {Mead}(1965)}]{nelder1965simplex}%
  \BibitemOpen
  \bibfield  {author} {\bibinfo {author} {\bibfnamefont {J.~A.}\ \bibnamefont
  {Nelder}}\ and\ \bibinfo {author} {\bibfnamefont {R.}~\bibnamefont {Mead}},\
  }\bibfield  {title} {\bibinfo {title} {A simplex method for function
  minimization},\ }\href@noop {} {\bibfield  {journal} {\bibinfo  {journal}
  {The computer journal}\ }\textbf {\bibinfo {volume} {7}},\ \bibinfo {pages}
  {308} (\bibinfo {year} {1965})}\BibitemShut {NoStop}%
\bibitem [{\citenamefont {Zhou}\ \emph {et~al.}(2020)\citenamefont {Zhou},
  \citenamefont {Wang}, \citenamefont {Choi}, \citenamefont {Pichler},\ and\
  \citenamefont {Lukin}}]{zhou2020quantum}%
  \BibitemOpen
  \bibfield  {author} {\bibinfo {author} {\bibfnamefont {L.}~\bibnamefont
  {Zhou}}, \bibinfo {author} {\bibfnamefont {S.-T.}\ \bibnamefont {Wang}},
  \bibinfo {author} {\bibfnamefont {S.}~\bibnamefont {Choi}}, \bibinfo {author}
  {\bibfnamefont {H.}~\bibnamefont {Pichler}},\ and\ \bibinfo {author}
  {\bibfnamefont {M.~D.}\ \bibnamefont {Lukin}},\ }\bibfield  {title} {\bibinfo
  {title} {Quantum approximate optimization algorithm: Performance, mechanism,
  and implementation on near-term devices},\ }\href@noop {} {\bibfield
  {journal} {\bibinfo  {journal} {Physical Review X}\ }\textbf {\bibinfo
  {volume} {10}},\ \bibinfo {pages} {021067} (\bibinfo {year}
  {2020})}\BibitemShut {NoStop}%
\bibitem [{\citenamefont {Wiersema}\ \emph {et~al.}(2020)\citenamefont
  {Wiersema}, \citenamefont {Zhou}, \citenamefont {de~Sereville}, \citenamefont
  {Carrasquilla}, \citenamefont {Kim},\ and\ \citenamefont
  {Yuen}}]{wiersema2020exploring}%
  \BibitemOpen
  \bibfield  {author} {\bibinfo {author} {\bibfnamefont {R.}~\bibnamefont
  {Wiersema}}, \bibinfo {author} {\bibfnamefont {C.}~\bibnamefont {Zhou}},
  \bibinfo {author} {\bibfnamefont {Y.}~\bibnamefont {de~Sereville}}, \bibinfo
  {author} {\bibfnamefont {J.~F.}\ \bibnamefont {Carrasquilla}}, \bibinfo
  {author} {\bibfnamefont {Y.~B.}\ \bibnamefont {Kim}},\ and\ \bibinfo {author}
  {\bibfnamefont {H.}~\bibnamefont {Yuen}},\ }\bibfield  {title} {\bibinfo
  {title} {Exploring entanglement and optimization within the hamiltonian
  variational ansatz},\ }\href@noop {} {\bibfield  {journal} {\bibinfo
  {journal} {PRX Quantum}\ }\textbf {\bibinfo {volume} {1}},\ \bibinfo {pages}
  {020319} (\bibinfo {year} {2020})}\BibitemShut {NoStop}%
\bibitem [{\citenamefont {Skolik}\ \emph {et~al.}(2021)\citenamefont {Skolik},
  \citenamefont {McClean}, \citenamefont {Mohseni}, \citenamefont {van~der
  Smagt},\ and\ \citenamefont {Leib}}]{skolik2021layerwise}%
  \BibitemOpen
  \bibfield  {author} {\bibinfo {author} {\bibfnamefont {A.}~\bibnamefont
  {Skolik}}, \bibinfo {author} {\bibfnamefont {J.~R.}\ \bibnamefont {McClean}},
  \bibinfo {author} {\bibfnamefont {M.}~\bibnamefont {Mohseni}}, \bibinfo
  {author} {\bibfnamefont {P.}~\bibnamefont {van~der Smagt}},\ and\ \bibinfo
  {author} {\bibfnamefont {M.}~\bibnamefont {Leib}},\ }\bibfield  {title}
  {\bibinfo {title} {Layerwise learning for quantum neural networks},\
  }\href@noop {} {\bibfield  {journal} {\bibinfo  {journal} {Quantum Machine
  Intelligence}\ }\textbf {\bibinfo {volume} {3}},\ \bibinfo {pages} {1}
  (\bibinfo {year} {2021})}\BibitemShut {NoStop}%
\bibitem [{\citenamefont {Campos}\ \emph {et~al.}(2021)\citenamefont {Campos},
  \citenamefont {Nasrallah},\ and\ \citenamefont
  {Biamonte}}]{campos2021abrupt}%
  \BibitemOpen
  \bibfield  {author} {\bibinfo {author} {\bibfnamefont {E.}~\bibnamefont
  {Campos}}, \bibinfo {author} {\bibfnamefont {A.}~\bibnamefont {Nasrallah}},\
  and\ \bibinfo {author} {\bibfnamefont {J.}~\bibnamefont {Biamonte}},\
  }\bibfield  {title} {\bibinfo {title} {Abrupt transitions in variational
  quantum circuit training},\ }\href@noop {} {\bibfield  {journal} {\bibinfo
  {journal} {Physical Review A}\ }\textbf {\bibinfo {volume} {103}},\ \bibinfo
  {pages} {032607} (\bibinfo {year} {2021})}\BibitemShut {NoStop}%
\bibitem [{\citenamefont {Hamilton}\ \emph {et~al.}(2019)\citenamefont
  {Hamilton}, \citenamefont {Dumitrescu},\ and\ \citenamefont
  {Pooser}}]{hamilton2019generative}%
  \BibitemOpen
  \bibfield  {author} {\bibinfo {author} {\bibfnamefont {K.~E.}\ \bibnamefont
  {Hamilton}}, \bibinfo {author} {\bibfnamefont {E.~F.}\ \bibnamefont
  {Dumitrescu}},\ and\ \bibinfo {author} {\bibfnamefont {R.~C.}\ \bibnamefont
  {Pooser}},\ }\bibfield  {title} {\bibinfo {title} {Generative model
  benchmarks for superconducting qubits},\ }\href@noop {} {\bibfield  {journal}
  {\bibinfo  {journal} {Physical Review A}\ }\textbf {\bibinfo {volume} {99}},\
  \bibinfo {pages} {062323} (\bibinfo {year} {2019})}\BibitemShut {NoStop}%
\bibitem [{\citenamefont {MindFoundary}(2021)}]{OPTaaS}%
  \BibitemOpen
  \bibfield  {author} {\bibinfo {author} {\bibnamefont {MindFoundary}},\ }\href
  {https://optaas.mindfoundry.ai/static/swagger/index.html} {\bibinfo {title}
  {Mindfoundry bayesian optimizer}} (\bibinfo {year} {2021})\BibitemShut
  {NoStop}%
\bibitem [{\citenamefont {Collins}(2003)}]{8178732}%
  \BibitemOpen
  \bibfield  {author} {\bibinfo {author} {\bibfnamefont {B.}~\bibnamefont
  {Collins}},\ }\bibfield  {title} {\bibinfo {title} {Moments and cumulants of
  polynomial random variables on unitarygroups, the itzykson-zuber integral,
  and free probability},\ }\href {https://doi.org/10.1155/S107379280320917X}
  {\bibfield  {journal} {\bibinfo  {journal} {International Mathematics
  Research Notices}\ }\textbf {\bibinfo {volume} {2003}},\ \bibinfo {pages}
  {953} (\bibinfo {year} {2003})}\BibitemShut {NoStop}%
\bibitem [{\citenamefont {Pucha{\l}a}\ and\ \citenamefont
  {Miszczak}(2011)}]{puchala2011symbolic}%
  \BibitemOpen
  \bibfield  {author} {\bibinfo {author} {\bibfnamefont {Z.}~\bibnamefont
  {Pucha{\l}a}}\ and\ \bibinfo {author} {\bibfnamefont {J.~A.}\ \bibnamefont
  {Miszczak}},\ }\bibfield  {title} {\bibinfo {title} {Symbolic integration
  with respect to the haar measure on the unitary group},\ }\href@noop {}
  {\bibfield  {journal} {\bibinfo  {journal} {arXiv preprint arXiv:1109.4244}\
  } (\bibinfo {year} {2011})}\BibitemShut {NoStop}%
\bibitem [{\citenamefont {Brandao}\ \emph {et~al.}(2016)\citenamefont
  {Brandao}, \citenamefont {Harrow},\ and\ \citenamefont
  {Horodecki}}]{brandao2016local}%
  \BibitemOpen
  \bibfield  {author} {\bibinfo {author} {\bibfnamefont {F.~G.}\ \bibnamefont
  {Brandao}}, \bibinfo {author} {\bibfnamefont {A.~W.}\ \bibnamefont
  {Harrow}},\ and\ \bibinfo {author} {\bibfnamefont {M.}~\bibnamefont
  {Horodecki}},\ }\bibfield  {title} {\bibinfo {title} {Local random quantum
  circuits are approximate polynomial-designs},\ }\href@noop {} {\bibfield
  {journal} {\bibinfo  {journal} {Communications in Mathematical Physics}\
  }\textbf {\bibinfo {volume} {346}},\ \bibinfo {pages} {397} (\bibinfo {year}
  {2016})}\BibitemShut {NoStop}%
\bibitem [{\citenamefont {Harrow}\ and\ \citenamefont
  {Mehraban}(2018)}]{harrow2018approximate}%
  \BibitemOpen
  \bibfield  {author} {\bibinfo {author} {\bibfnamefont {A.}~\bibnamefont
  {Harrow}}\ and\ \bibinfo {author} {\bibfnamefont {S.}~\bibnamefont
  {Mehraban}},\ }\bibfield  {title} {\bibinfo {title} {Approximate unitary $ t
  $-designs by short random quantum circuits using nearest-neighbor and
  long-range gates},\ }\href@noop {} {\bibfield  {journal} {\bibinfo  {journal}
  {arXiv preprint arXiv:1809.06957}\ } (\bibinfo {year} {2018})}\BibitemShut
  {NoStop}%
\bibitem [{\citenamefont {Roth}\ \emph {et~al.}(2018)\citenamefont {Roth},
  \citenamefont {Kueng}, \citenamefont {Kimmel}, \citenamefont {Liu},
  \citenamefont {Gross}, \citenamefont {Eisert},\ and\ \citenamefont
  {Kliesch}}]{roth2018recovering}%
  \BibitemOpen
  \bibfield  {author} {\bibinfo {author} {\bibfnamefont {I.}~\bibnamefont
  {Roth}}, \bibinfo {author} {\bibfnamefont {R.}~\bibnamefont {Kueng}},
  \bibinfo {author} {\bibfnamefont {S.}~\bibnamefont {Kimmel}}, \bibinfo
  {author} {\bibfnamefont {Y.-K.}\ \bibnamefont {Liu}}, \bibinfo {author}
  {\bibfnamefont {D.}~\bibnamefont {Gross}}, \bibinfo {author} {\bibfnamefont
  {J.}~\bibnamefont {Eisert}},\ and\ \bibinfo {author} {\bibfnamefont
  {M.}~\bibnamefont {Kliesch}},\ }\bibfield  {title} {\bibinfo {title}
  {Recovering quantum gates from few average gate fidelities},\ }\href@noop {}
  {\bibfield  {journal} {\bibinfo  {journal} {Physical review letters}\
  }\textbf {\bibinfo {volume} {121}},\ \bibinfo {pages} {170502} (\bibinfo
  {year} {2018})}\BibitemShut {NoStop}%
\bibitem [{\citenamefont {Preskill}(2018)}]{preskill2018quantum}%
  \BibitemOpen
  \bibfield  {author} {\bibinfo {author} {\bibfnamefont {J.}~\bibnamefont
  {Preskill}},\ }\bibfield  {title} {\bibinfo {title} {Quantum computing in the
  nisq era and beyond},\ }\href@noop {} {\bibfield  {journal} {\bibinfo
  {journal} {Quantum}\ }\textbf {\bibinfo {volume} {2}},\ \bibinfo {pages} {79}
  (\bibinfo {year} {2018})}\BibitemShut {NoStop}%
\bibitem [{\citenamefont {Arrasmith}\ \emph {et~al.}(2020)\citenamefont
  {Arrasmith}, \citenamefont {Cerezo}, \citenamefont {Czarnik}, \citenamefont
  {Cincio},\ and\ \citenamefont {Coles}}]{arrasmith2020effect}%
  \BibitemOpen
  \bibfield  {author} {\bibinfo {author} {\bibfnamefont {A.}~\bibnamefont
  {Arrasmith}}, \bibinfo {author} {\bibfnamefont {M.}~\bibnamefont {Cerezo}},
  \bibinfo {author} {\bibfnamefont {P.}~\bibnamefont {Czarnik}}, \bibinfo
  {author} {\bibfnamefont {L.}~\bibnamefont {Cincio}},\ and\ \bibinfo {author}
  {\bibfnamefont {P.~J.}\ \bibnamefont {Coles}},\ }\bibfield  {title} {\bibinfo
  {title} {Effect of barren plateaus on gradient-free optimization},\
  }\href@noop {} {\bibfield  {journal} {\bibinfo  {journal} {arXiv preprint
  arXiv:2011.12245}\ } (\bibinfo {year} {2020})}\BibitemShut {NoStop}%
\bibitem [{\citenamefont {Anschuetz}(2021)}]{anschuetz2021critical}%
  \BibitemOpen
  \bibfield  {author} {\bibinfo {author} {\bibfnamefont {E.~R.}\ \bibnamefont
  {Anschuetz}},\ }\bibfield  {title} {\bibinfo {title} {Critical points in
  hamiltonian agnostic variational quantum algorithms},\ }\href@noop {}
  {\bibfield  {journal} {\bibinfo  {journal} {arXiv preprint arXiv:2109.06957}\
  } (\bibinfo {year} {2021})}\BibitemShut {NoStop}%
\bibitem [{\citenamefont {Arrasmith}\ \emph {et~al.}(2019)\citenamefont
  {Arrasmith}, \citenamefont {Cincio}, \citenamefont {Sornborger},
  \citenamefont {Zurek},\ and\ \citenamefont
  {Coles}}]{arrasmith2019variational}%
  \BibitemOpen
  \bibfield  {author} {\bibinfo {author} {\bibfnamefont {A.}~\bibnamefont
  {Arrasmith}}, \bibinfo {author} {\bibfnamefont {L.}~\bibnamefont {Cincio}},
  \bibinfo {author} {\bibfnamefont {A.~T.}\ \bibnamefont {Sornborger}},
  \bibinfo {author} {\bibfnamefont {W.~H.}\ \bibnamefont {Zurek}},\ and\
  \bibinfo {author} {\bibfnamefont {P.~J.}\ \bibnamefont {Coles}},\ }\bibfield
  {title} {\bibinfo {title} {Variational consistent histories as a hybrid
  algorithm for quantum foundations},\ }\href@noop {} {\bibfield  {journal}
  {\bibinfo  {journal} {Nature communications}\ }\textbf {\bibinfo {volume}
  {10}},\ \bibinfo {pages} {1} (\bibinfo {year} {2019})}\BibitemShut {NoStop}%
\bibitem [{\citenamefont {Fletcher}(2013)}]{fletcher2013practical}%
  \BibitemOpen
  \bibfield  {author} {\bibinfo {author} {\bibfnamefont {R.}~\bibnamefont
  {Fletcher}},\ }\href@noop {} {\emph {\bibinfo {title} {Practical methods of
  optimization}}}\ (\bibinfo  {publisher} {John Wiley \& Sons},\ \bibinfo
  {year} {2013})\BibitemShut {NoStop}%
\bibitem [{\citenamefont {Cerezo}\ \emph {et~al.}(2020)\citenamefont {Cerezo},
  \citenamefont {Sone}, \citenamefont {Volkoff}, \citenamefont {Cincio},\ and\
  \citenamefont {Coles}}]{cerezo2020cost}%
  \BibitemOpen
  \bibfield  {author} {\bibinfo {author} {\bibfnamefont {M.}~\bibnamefont
  {Cerezo}}, \bibinfo {author} {\bibfnamefont {A.}~\bibnamefont {Sone}},
  \bibinfo {author} {\bibfnamefont {T.}~\bibnamefont {Volkoff}}, \bibinfo
  {author} {\bibfnamefont {L.}~\bibnamefont {Cincio}},\ and\ \bibinfo {author}
  {\bibfnamefont {P.~J.}\ \bibnamefont {Coles}},\ }\bibfield  {title} {\bibinfo
  {title} {Cost-function-dependent barren plateaus in shallow quantum neural
  networks},\ }\href@noop {} {\bibfield  {journal} {\bibinfo  {journal} {arXiv
  e-prints}\ ,\ \bibinfo {pages} {arXiv}} (\bibinfo {year} {2020})}\BibitemShut
  {NoStop}%
\end{thebibliography}%

\end{document}